\documentclass[aps,pra,superscriptaddress,twocolumn]{revtex4-2}
\usepackage{amsmath}
\usepackage{amsfonts}
\usepackage{graphicx}
\usepackage{xcolor}

\usepackage{braket} 

\begin{document}

\title{Parametric model for high-order harmonic generation with quantized fields}
\author{\'{A}kos Gombk\"{o}t\H{o}}
\affiliation{Wigner Research Centre for Physics, Konkoly-Thege M. \'{u}t 29-33, H-1121 Budapest, Hungary}
\affiliation{University of Pannonia, Zalaegerszeg Campus, Gasparich M\'ark u. 18/A, H-8900 Zalaegerszeg, Hungary}
\author{S\'{a}ndor Varr\'{o}}
\affiliation{ELI-ALPS, ELI-HU Non-profit Ltd., Dugonics t\'{e}r 13, H-6720 Szeged, Hungary}%
\author{B\'{e}la G\'{a}bor Pusztai}
\affiliation{Bolyai Institute, University of Szeged, Aradi vértanúk tere 1, H-6720 Szeged, Hungary}%
\affiliation{Department of Basic Sciences, Faculty of Mechanical Engineering and Automation, John von Neumann University, Izs\'aki \'ut 10, H-6000 Kecskem\'et, Hungary}
\author{Istv\'{a}n Magashegyi}
\affiliation{Department of Theoretical Physics, University of Szeged, Tisza Lajos k\"{o}r%
\'{u}t 84, H-6720 Szeged, Hungary}
\affiliation{Wigner Research Centre for Physics, Konkoly-Thege M. \'{u}t 29-33, H-1121 Budapest, Hungary}
\author{Attila Czirj\'ak}
\affiliation{Department of Theoretical Physics, University of Szeged, Tisza Lajos k\"{o}r%
\'{u}t 84, H-6720 Szeged, Hungary}
\affiliation{ELI-ALPS, ELI-HU Non-profit Ltd., Dugonics t\'{e}r 13, H-6720 Szeged, Hungary}%
\author{Szabolcs Hack}
\affiliation{Department of Theoretical Physics, University of Szeged, Tisza Lajos k\"{o}r%
\'{u}t 84, H-6720 Szeged, Hungary}
\affiliation{ELI-ALPS, ELI-HU Non-profit Ltd., Dugonics t\'{e}r 13, H-6720 Szeged, Hungary}%
\author{P\'{e}ter F\"{o}ldi}
\affiliation{Department of Theoretical Physics, University of Szeged, Tisza Lajos k\"{o}r%
\'{u}t 84, H-6720 Szeged, Hungary}
\affiliation{ELI-ALPS, ELI-HU Non-profit Ltd., Dugonics t\'{e}r 13, H-6720 Szeged, Hungary}%

\begin{abstract}
A quantum optical model for the high-order harmonic generation is presented, in which both the exciting field and the high harmonic modes are quantized, while the target material appears via parameters only. As a consequence, the model is independent from the excited material system to a large extent, and allows us to focus on the properties of the electromagnetic fields. Technically, the Hamiltonian known for parametric down-conversion is adopted, where photons in the $n$th harmonic mode are created in exchange of annihilating $n$ photons from the fundamental mode. In our treatment, initially the fundamental mode is in a coherent state corresponding to large photon numbers, while the high harmonic modes are in vacuum state. Due to the interaction, the latter modes get populated while the fundamental one loses photons. Analytical approximations are presented for the time evolution that are verified by numerically exact calculations. For multimode, finite bandwith
excitation, the time dependence of the high-order harmonic radiation is also given.
\end{abstract}

\maketitle

\section{Introduction}
Optical generation of high-order harmonics (HH) \cite{McPherson87,Ferray88} is a key process for the generation of ultrashort electromagnetic pulses \cite{FT1992}. It is a strong-field phenomenon, in which the nonlinear material response leads to the emission of radiation with frequencies close to the integer multiples of the central frequency of the excitation. The effect has been demonstrated using various material samples \cite{McPherson87,Ferray88,QTMDMGA2006,Vincenti2014,GDiCSADiMR11,Ghim19}, and although the details are different, the exciting field is always orders of magnitude stronger than the harmonics. Therefore, besides the traditional models \cite{Cor93,Lew94} for high-order harmonic generation (HHG), quantized description (photon picture) is also of interest. Interestingly, observable signatures of the back-action of the HH modes on the photon statistics of the fundamental one have recently been demonstrated experimentally \cite{Tsatrafyllis2017,T19}. This underlines the significance of field quantization during the process of HHG.

Considering the quantized description of strong-field effects, early theoretical models date back to the '80s \cite{BV81b,BV81a}. In this framework, Ref.~\cite{BV81b} provides the first non-perturbative treatment of HHG in the nonlinear Compton process. For HHG with atomic samples, Volkov states and transitions between them have also been considered \cite{GSE98,CCL00}. More recently, theoretical models in Refs.~\cite{Gonos16,T19} provide a background for the corresponding experimental findings reported in Refs.~\cite{Tsatrafyllis2017,T19}. The emergence of HHG spectra as the expectation value of the photon numbers in the HH modes were demonstrated in \cite{GCVF16,AGPF21,photonics8070263,GA23}, and a phase space description of the problem was given in Ref.~\cite{GVMF20}. Explicitly nonclassical features of the HH emission have been predicted also for realistic systems \cite{Gorlach20}, and the consequences of excitation with quantum light have also been investigated \cite{Gorlach23}.  For a recent tutorial on strong-field quantum electrodynamics, see Ref.~\cite{LewPRX23}.

The complete problem, i.e., considering both the excitation and the HH modes on the quantized level (with the material sample obviously treated quantum mechanically), raises numerical challenges. With reasonable approximations, either the HH modes \cite{GCVF16}, or the excitation \cite{GVMF20} can be considered to be quantized, but it is demanding to solve a model, which assumes it for both fields. In order to focus on the most important quantum mechanical properties both of these fields, we have to use a model in which the number of different degrees of freedom is systematically decreased. This is indeed the aim of the current paper, in which, in a rather general way, the target material appears only via parameters -- that can be fitted to the given material system in subsequent, more specific calculations.

\section{Model}
The essence of high-order harmonic generation on the level of quantized fields is the creation of photons in the HH modes at the expense of decreasing the number of photons in the exciting (fundamental) modes. Clearly, in  order to obtain a Hermitian Hamiltonian, we have to take the opposite process also into account. That is, by denoting the fundamental frequency by $\omega$, we can write
\begin{equation}
\begin{aligned}
H = &\hbar \omega \left(A^\dagger A +\frac{1}{2}\right)+ \hbar \omega \sum_n n \left(a_n^\dagger a_n +\frac{1}{2}\right) \\
+& \hbar \sum_n \chi_n \left[ a_n^\dagger A^n +a_n (A^\dagger)^n\right],
\label{Ham}
\end{aligned}
\end{equation}
where $a_n$ $(a_n^\dagger)$ are the annihilation (creation) operator of a mode with frequency $n\omega,$ while $A$ and $A^\dagger$ belong to the fundamental mode. The coefficients $\chi_n$ are real numbers. The magnitude of these parameters is to be determined so as to reproduce experimental results, see in the next section and the Appendix.

\bigskip

The model proposed in Eq.~(\ref{Ham}) is clearly a simplification from various points of view. Obviously, the excitation is generally multi-mode, and has a finite duration. Similarly, the spectral widths of the high-order harmonics are also finite, and the HH peaks may not be positioned exactly at the integer multiples of $\omega.$ (Specifically, for most samples, harmonics corresponding to even values of $n$ are missing, as we shall assume in the following.) Moreover, the dynamics of the electromagnetic modes (which can be obtained by "tracing out" the degrees of freedom corresponding to the target material) can be more complex than the unitary time evolution induced by the Hamiltonian (\ref{Ham}).

However, Eq.~(\ref{Ham}) -- besides being instructive -- clearly captures the most important features of the HHG process. Additionally, when only a single harmonic corresponding to $n=2$ is taken into account, (\ref{Ham}) is identical to the Hamiltonian that is used for the description of parametric down-conversion (when the process $2\omega \rightarrow \omega$ is of interest \cite{WT71}). That is, Eq.~(\ref{Ham}), with appropriate initial conditions, is closely related to the description of fundamental experiments related to photon pair creation \cite{Kwiat95,Z22}.

The natural initial condition for HHG is that the state of the fundamental mode (the mode that corresponds to an intense, practically classical radiation) is a coherent state, while the HH modes are in vacuum state:
\begin{equation}\label{initial}
  \vert \Psi\rangle = \vert \alpha_{0}\rangle \otimes \vert 0 \rangle_{n_1} \otimes \vert 0 \rangle_{n_2} \otimes \cdots \otimes \vert 0 \rangle_N,
\end{equation}
where $N$ denotes the order of the highest harmonic that we take into account. In the following all the indices $n_i$ are considered to be odd, and we focus on the plateau region where the heights of the harmonic peaks are practically the same. This region is present for both gaseous and solid state target materials. In the power spectra, peaks that correspond to different harmonic frequencies and have equal heights mean the same amount of energy irradiated in the narrow spectral regions around the different frequencies $n_i \omega.$  In order to reproduce this effect in our model, the expectation values $\hbar n_i \omega \langle a^\dagger_{n_i} a_{n_i}\rangle$ should be independent of $n_i$. This can be achieved by choosing appropriate ratios of the parameters $\chi_i,$ as we shall see in the next section.

\section{Parametric approximation}
\label{Paramsec}
Since the intensity of the HH radiation is known to be orders of magnitude lower than that of the excitation, we can assume that the back-action of the HH modes on the fundamental one is weak. Therefore, as a first approximation, we can take the expectation value of Eq.~(\ref{Ham}) in the state $|\alpha (t)\rangle=|\alpha_0 \exp(-i \omega t)\rangle,$ which would be the result of the free time evolution of the fundamental mode [induced by the Hamiltonian $\hbar \omega \left(A^\dagger A \right)$]. Using $\langle \alpha (t)|\alpha(t)\rangle=1$ and the eigenvalue equation $A|\alpha(t)\rangle=\alpha(t) |\alpha(t)\rangle,$ we obtain the following Hamiltonian in the parametric approximation:
\begin{equation}
H_p =  \sum_n \hbar \omega  n \left(a_n^\dagger a_n \right) +  \hbar \chi_n \left[ a_n^\dagger \alpha_0^n e^{-i n \omega t} +a_n (\alpha_0^*)^n e^{i n \omega t} \right],
\label{Hamp}
\end{equation}
where irrelevant additive constants have been omitted. (Note that this approximation is common for describing down-conversion \cite{WM94}; its generalization to cases when the pump mode is not necessarily a coherent state can be found in Ref.~\cite{HDB94}.) As we can see, $H_p$ is a sum of independent terms, each of which can be treated separately. Thus, dropping the index $n$ for the moment, we are to consider the single mode Hamiltonian
\begin{equation}
H_{p,s} =  \hbar \Omega  \left(a^\dagger a \right) +  \hbar \left[ a^\dagger \beta(t) +a \beta^*(t)\right],
\label{Hampn}
\end{equation}
where $\Omega=n \omega$ and $\beta(t)=\chi_n \alpha_0^n e^{-i n \omega t}.$

The time evolution induced by (\ref{Hampn}) can be solved using a systematic, standard way \cite{L61,G63}. However, it can be instructive to perform a direct calculation. To this end, we use the Heisenberg picture.
The commutator $[a,a^\dagger]=1$ leads to the following dynamical equation for the annihilation operator:
\begin{equation}
i \hbar \dot{a}_H =  \hbar \Omega a_H +  \hbar \beta(t) \mathbf{I},
\label{dota}
\end{equation}
where $\mathbf{I}$ denotes the identity operator, and the index $H$ refers to the Heisenberg picture. This equation can be solved:
\begin{equation}
a_H(t) = a_H(0) e^{-i \Omega t} - i t \beta(t) \mathbf{I}.
\label{dota}
\end{equation}
Now let us use the time evolution operator, $U,$ for which, with any initial state $|\phi(0)\rangle,$ we have $U(t)|\phi(0)\rangle=|\phi(t)\rangle.$ Clearly, $a=a_S=U(t)a_H(t)U^\dagger(t),$ where the explicit notation of the Schr\"odinger picture (the index $S$) has been introduced for clarity. Since $a_H(0)|0\rangle =0,$ we have
\begin{equation}
a_H(t)|0\rangle = -i t \beta(t)|0\rangle.
\label{Heis}
\end{equation}
Returning to the Schr\"odinger picture, and using that $U(t)U^\dagger(t)=\mathbf{I},$ we can write
\begin{equation}
U(t)a_H(t)U^\dagger(t)U(t)|0\rangle = -i t \beta(t)U(t)|0\rangle.
\label{HS}
\end{equation}
That is, the time evolution starting from the vacuum state $|0\rangle$ (which is the initial condition, see Eq.~(\ref{initial})) leads to a state that is an eigenstate of $a_S:$
\begin{equation}
a_S U(t)|0\rangle = -i t \beta(t)U(t)|0\rangle.
\end{equation}
This means that the state $U(t)|0\rangle$ is a coherent state with the parameter of $-i t \beta(t),$ that is
\begin{equation}
|\phi(t)\rangle=U(t)|0\rangle = |-i t \beta(t)\rangle.
\label{cohbeta}
\end{equation}
{Before analyzing this result, let us emphasize an important aspect that has experimental relevance as well. This is related to the phases of the various harmonics, which have to be locked relative to each other \cite{FT1992} in order that their superposition could lead to an isolated attosecond pulse, or a pulse train \cite{Krausz_Nobel,Agostini_Nobel}. Obviously, the phases of the coherent states (\ref{cohbeta}) corresponding to different harmonics are linked via the common exciting state $|\alpha_0\rangle,$ and consequently the same holds for the related electromagnetic field. Although at this point it can be seen as a consequence of the parametric approximation, our numerically exact calculations show that the relative phases of the different harmonics are practically constants at the initial stage of the time evolution.}

As Eq.~(\ref{cohbeta}) implies, the photon number expectation value in the parametric approximation is given by
\begin{equation}
\langle a^\dagger_S  a_S \rangle (t)=t^2 |\beta(t)|^2,
\label{Nexp}
\end{equation}
i.e., it is quadratic in $t$. Clearly, it cannot hold for infinitely long times, which is a direct consequence of the approximation that the photon number expectation value of the fundamental mode is constant \cite{H90}. As we shall see in the next section, for physically realistic interaction times, this approximation is numerically proven to be acceptable. Additionally, Eq.~(\ref{Nexp}) allows us to estimate the coefficients $\chi_n$ that lead to peaks in the power spectra with equal heights. The corresponding requirement is that $\hbar\Omega\langle a^\dagger a\rangle$ is the same for all HH modes (c.f.~the end of the previous section). Reintroducing the mode indices, and using Eq.~(\ref{Nexp}), we see that
\begin{equation}
\frac{\chi_n}{\chi_{n'}} = \frac{\sqrt{n'}|\alpha_0|^{n'}}{\sqrt{n} |\alpha_0|^{n}}
\label{plateau}
\end{equation}
describes the ratio between the $\chi$ parameters that correspond to two different HH modes ($n$ and $n'$) in the plateau region. In view of this, the parameter
\begin{equation}
p={\chi_n}{\sqrt{n} |\alpha_0|^{n}}
\label{Pparam}
\end{equation}
will be kept constant (independent of $n$) in the following in order to account for the plateau part of the spectrum. {(A different choice for the parameters $\chi_n$ that corresponds to actual experimental data will be given in Appendix B.)}

Note that the plateau condition that led to Eq.~(\ref{Pparam}) can be too strong in the sense that it is based on an approximation, and the exact time evolution can be slightly different (see the next section). Additionally, for pulsed excitation, the HH photons are collected during the whole process, and the condition of having HH peaks with equal heights is to be fulfilled only at the end of the process. Fitting the parameters to different target materials may also slightly modify the condition (\ref{Pparam}). However, at the general level of the current work, this is a reasonable approximation.

\bigskip

The results to follow will be presented using the Schr\"odinger picture, without explicit notations (there will be no index $S$ from now on). Before turning to the numerical results, let us consider the case of multimode excitation within the framework above. We assume a pulsed-like excitation and for the sake of simplicity we consider discrete frequencies. As before, we restrict ourselves to the linearly polarized case. This means that we assume a (large) quantization volume $V,$ in which the electromagnetic field with the given polarization direction can be written as a sum of contributions of modes with discrete frequencies. In Eq.~(\ref{Ham}) we took only a very limited number of modes into account, but in order to describe a pulsed-like excitation, even in free space, we need all the modes within the spectral range $[\omega_{min},\omega_{max}]$ of the excitation. Let us consider the multimode exciting coherent state
\begin{equation}
  \vert \mathbf{\boldsymbol{\alpha}}_0\rangle = \Pi_m^\otimes\vert \alpha_{0,m}\rangle,
\label{multicoh}
\end{equation}
where the frequencies corresponding to $\vert \alpha_{0,m}\rangle$ -- to be denoted by $\omega_m$ -- fall in the range $[\omega_{min},\omega_{max}]$. Knowing e.g., the experimentally available "waveform" $E(t)$ of the exciting electric field,  the complex parameters $\alpha_{0,m}$ can be chosen such that without any interaction, the expectation value of the quantum mechanical electric field operator, $\langle E \rangle(t)\propto i \sum_m \left [ \langle A_m \rangle(t)- \langle A_m^\dagger \rangle(t)\right ],$ is the same as $E(t).$

In more detail, considering $H_0=\hbar \omega A_m^\dagger A_m $ alone ("free space") we have $|\alpha_m\rangle(t)=|\alpha_{0,m}\exp(-i \omega_m t)\rangle,$ thus the expectation value $\langle A_m \rangle(t) $ is given by $\alpha_{0,m}\exp(-i \omega_m t)$, and consequently we are essentially to determine the Fourier components of $E(t)$ to obtain the required $\alpha_{0,m}$ values.  Alternatively, working in velocity gauge, the coefficients $\alpha_{0,m}$ can be chosen so that the time evolution of the classical vector potential is reproduced by its quantum mechanical expectation value, ensuring that it holds also for both the electric and the magnetic fields.

Let us note that the mode expansion is usually considerably more complex than the one given -- quite formally -- by Eq.~(\ref{multicoh}), even for a given polarization direction. E.g.~the transversal structure of the modes and the mode density should also be taken into account for a detailed description. However, in order to see the approximate time evolution of the HH electric fields, we can consider Eq.~(\ref{multicoh}) as the initial state for the modes corresponding to the exciting pulse. Using the free-space time evolution of the coherent states, we have:
\begin{equation}
\begin{aligned}
E(t)&=\langle E \rangle (t)=  i \sum_m \mathcal{E}(\omega_m) \left[\langle A_m\rangle (t) - \langle A_m^\dagger \rangle (t) \right] \\
&= i \sum_m \mathcal{E}(\omega_m) \left[\alpha_{0,m} e^{-i\omega_m t}- \alpha^*_{0,m} e^{i\omega_m t} \right],
\end{aligned}
\label{Etime}
\end{equation}
where the distribution of the frequencies $\omega_m$ are peaked around the fundamental one, i.e., $\omega,$ and $\mathcal{E}(\nu) =\sqrt{\frac{\hbar \nu}{2\epsilon_0 V}}$ (recall that $V$ denotes the quantization volume). Now we use a further approximation, namely we assume that each harmonic state in the expansion (\ref{multicoh}) populates the corresponding HH modes (i.e., $\omega_m \rightarrow n \omega_m$) independently. (Clearly, depending on the mode density, it can be a strong simplification, thus the following results are to be considered in the qualitative sense.) According to Eq.~(\ref{cohbeta}), for the $n$th harmonic we can write:
\begin{equation}
\begin{aligned}
\langle E_n \rangle (t) &=  i \sum_m \mathcal{E}(n \omega_{m}) \left[ \langle a_{n,m}\rangle (t)-\langle a_{n,m}^\dagger  \rangle (t) \right] \\
& = t \chi_n \sum_m \mathcal{E}(n\omega_m) \left [\alpha^n_{0,m} e^{-in \omega_m t} + c.c. \right],
\label{Etimen}
\end{aligned}
\end{equation}
where $a_{n,m}$ and $a_{n,m}^\dagger$ correspond to the mode with frequency $n\omega_m,$ and, for the sake of simplicity, $\chi_n$ was considered to be constant in the frequency range $[\omega_{min},\omega_{max}].$

It is instructive to consider an example. For a pulse with Gaussian envelope, $E(t)=E_0 \exp(-t^2/\tau^2)\sin \omega t,$ we have $ \alpha_{0,m} \propto \exp[-\tau^2/4 (\omega+\omega_m)^2].$ Since $\mathcal{E}(n\omega_m)/\mathcal{E}(\omega_m)=\sqrt{n},$ independently from $m,$ by using Eq.~(\ref{Etimen}), we obtain that the time dependence of $\langle E_n \rangle$ is proportional to $t \exp(-n t^2/\tau^2)\cos (n \omega t),$ where the change of the carrier-envelope phase (CEP) ($\sin(\omega t)\rightarrow \cos(n \omega t)$) is a result of the factor $i$ in the coherent state $|-i t \beta(t)\rangle$.

When $\langle E_n \rangle (t)$ is compared to $\langle E \rangle (t),$ we see that there are obvious changes like replacing $\omega$ with $n \omega$  and the $n$ times decrease of the pulse duration. The change of the CEP and the multiplicative factor of $t$ are, however, less trivial. It is important to emphasize that $\langle E_n \rangle (t)$ does not diverge in the long time limit, as a consequence of the presence of the Gaussian envelope, $\langle E_n \rangle (t)\rightarrow 0$ as $t \rightarrow \infty.$   That is, in the framework of the parametric approximation, a pulsed-like excitation leads to pulsed-like HH radiation, but the waveforms corresponding to the HH modes are not simply scaled copies of the exciting pulse.

\section{Numerical results}

It convenient to rewrite the time dependent Schr\"odinger equation using dimensionless time $\tau=\omega t$ as
\begin{equation}\label{TDSE}
  i\frac{d}{d \tau}\vert \Psi(t) \rangle=\frac{1}{\hbar \omega} H\vert \Psi(t) \rangle,
\end{equation}
where the Hamiltonian $H$ is given by Eq.~(\ref{Ham}), and no additional approximations are used. Eq.~(\ref{TDSE}) tells us that the independent parameters are $\chi_n/\omega.$ Additionally, since the "weighted photon number operator"
\begin{equation}\label{Nop}
  N=A^\dagger A +\sum_n n a_n^\dagger a_n
\end{equation}
commutes with $H,$ the problem is finite dimensional. More precisely, working in the photon number eigenbasis $\vert n_0, n_1, \ldots n_N\rangle$ (for which $A^\dagger A \vert n_0, n_1, \ldots n_N\rangle=n_0 \vert n_0, n_1, \ldots n_N\rangle$ and $a_k^\dagger a_k \vert n_0, n_1, \ldots n_N\rangle=n_k \vert n_0, n_1, \ldots n_N\rangle$), we see that for an initial state $\vert \Psi(0) \rangle=\vert M,0,0,\ldots 0\rangle,$ $M$ is the maximal index for the fundamental mode, and the photon numbers cannot exceed $M/n$ for the $n$th harmonic mode either. As a consequence, for the physically interesting initial state given by Eq.~(\ref{initial}), the only truncation in photon number eigenbasis is related to the representation of the initial coherent state $\vert \alpha_0\rangle.$ (For nonzero $\alpha_0,$ the inner product $\langle n\vert \alpha_0\rangle$ is never zero exactly, but the photon number distribution is peaked around $|\alpha_0|^2$ (see e.g.~\cite{WM94}) and the state can be estimated using a finite number of photon eigenstates with arbitrary precision.)

\begin{figure}
\includegraphics[width=8.4cm]{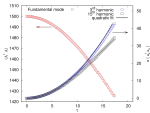}
\caption{The expectation values of the "weighted photon number operators", i.e., $A^\dagger A$ for the exciting (fundamental) mode (red circles, left vertical axis) and $n a_n^\dagger a_n$ for two HH modes ($n=3:$ blue open circles, $n=15:$ black open squares.) The solid black line is a quadratic fit for the $n=3$ case using the interval $\tau=[0,4].$ Parameters: $\alpha_0=\sqrt{1500}$, $p=0.2.$\label{Nexpini}}
\end{figure}

However, the difficulty of the numerical problem rapidly increases as we increase the number of modes that we take into account. The compromise between computation time and the requirement of accounting for different HH modes led us to consider two HH modes at a time, which allows us to take a few thousand exciting photons into account (i.e., $|\alpha_0|\approx \sqrt{1000}$).

\bigskip
Clearly, the photon number operators $A^\dagger A$ and $a_n^\dagger a_n$ commute with the interaction-free Hamiltonian of the system $H_0=\hbar\omega N$. Therefore, when calculating the time evolution of the photon number expectation values, it is only the interaction term
\begin{equation}\label{H_I}
  H_I=\hbar \sum_n \chi_n \left[ a_n^\dagger A^n +a_n (A^\dagger)^n\right]
\end{equation}
that has to be considered, e.g.:
\begin{equation}\label{H_I}
  i\frac{d}{d \tau} \langle A^\dagger A\rangle= \frac{1}{\hbar \omega} \langle \left[A^\dagger A, H_I \right]\rangle.
\end{equation}
This means that it is only the \emph{relative} magnitudes of the parameters that are essential. (In particular, if we take only a single HH mode ($n_1$) into account, the change $\chi_{n_1}\rightarrow \gamma \chi_{n_1}$ does not change the time evolution of $\langle A^\dagger A\rangle$ in a fundamental way, it only rescales the time variable $\tau\rightarrow \tau/\gamma.$) Since -- in order to reproduce the plateau region -- the ratio of parameters $\chi_n$ are fixed [see Eq.~(\ref{plateau})], the same scaling property applies to the general case. However, in order to be specific, in the following we will use $\tau,$ that is, the unit of time will be $1/\omega=T/2\pi$ (with $T$ denoting the exciting mode optical period). The parameter $p$ that is defined by Eq.~(\ref{Pparam}) will also be given for the figures.

\begin{figure}
\includegraphics[width=8.4cm]{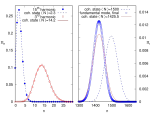}
\caption{The photon statistics of the modes indicated by the legend at the end of the time evolution shown by Fig.~\ref{Nexpini} $ (\tau=18).$ The parameters are the same as in Fig.~\ref{Nexpini}. The left panel shows the HH modes, while the right one considers the exciting mode, where the photon statistics of the initial state ($\alpha_0=\sqrt{1500}$) is also shown. Circles denote numerical results, lines corresponding to coherent states serve as references. \label{Ndistr}}
\end{figure}

\subsection{The initial stage of the time evolution}
In the following, we present results obtained by solving the time dependent Schr\"odinger equation (\ref{TDSE}) numerically. The initial condition is given by Eq.~(\ref{initial}), and we consider two HH modes. In order to see the behaviour of significantly different modes, we chose $n_1=3$ and $n_2=15.$ The ratio $\chi_3/\chi_{15}$ is determined by Eq.~(\ref{plateau}). In this subsection we focus on the initial stage of the time evolution, in which (according to the previous section) the condition Eq.~(\ref{plateau}) should warrant that $3\langle a_3^\dagger a_3\rangle(t)=15\langle a_{15}^\dagger a_{15}\rangle$ within a good approximation. As we can see in Fig.~\ref{Nexpini}, indeed, this condition holds for small values of $\tau,$ and even at the end of the considered time interval, the relative difference is below 20 \%. {Clearly, this difference is a consequence of the decreasing number of photons in the exciting mode, which corresponds to the gradual loss of the validity of the parametric approximation. Since the parameter of the coherent state corresponding to the $n$th harmonic contains the $n$th power of $\alpha_0$, higher orders are more sensitive to this effect.}

The decrease of the exciting intensity can also be seen in Fig.~\ref{Ndistr}, where the numerically exact probability $\Pi_n$ of having $n$ photons in a given mode (the photon statistics) is shown. As we can also see, the photon statistics are close to that of coherent states, especially considering the HH modes. (The deviation from the Poissonian statistics of a coherent state can be quantitatively analyzed using the Mandel parameter, see the next subsection.)
More precisely, the probability $\Pi_n,$ e.g.~for the fundamental mode is given by
\begin{equation}\label{H_I}
  \Pi_n^{(0)}(t)=\langle \Psi (t) \vert n\rangle\langle n\vert \otimes  \mathbf{I} \otimes \cdots \otimes \mathbf{I}\vert\Psi(t)\rangle.
\end{equation}
At this point it is not necessary, but later it will be useful to introduce the reduced density matrices. For the $n$th mode, it is the trace of the projector $\vert\Psi\rangle\langle\Psi\vert$ over the modes different from $n.$ E.g., for $n=0:$
\begin{equation}\label{H_I}
  \rho^{(0)}(t)=\sum_{n_i, i\neq0} \langle n_1, n_2, \ldots n_N \vert \Psi(t)\rangle\langle \Psi(t)\vert  n_1, n_2, \ldots n_N\rangle.
\end{equation}
Using this notation, $\Pi_n^{(0)}(t)=\mathrm{Tr} \left ( \rho^{(0)}(t) \vert n \rangle \langle n \vert \right ).$

\begin{figure}
\includegraphics[width=8.4cm]{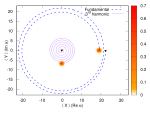}
\caption{The initial stage of the time evolution in phase space. Lines correspond to the expectation values of the quadrature operators $X$ and $Y$ for the fundamental mode (inward orienting spiral), and the third harmonic mode (spiral with increasing amplitude in accord with Eq.~(\ref{cohbeta})). The initial points of the trajectories ($\tau=0$) are denoted by black triangles. The Wigner functions corresponding to the final states {(at $\tau=4\pi$)} are also shown. (Note that for a coherent state $\vert \alpha\rangle,$ the expectation value of $X$ and $Y$ are $\mathrm{Re}\ \alpha$ and $\mathrm{Im}\ \alpha,$ respectively.) \label{Wshort}}
\end{figure}

\bigskip

Focusing on a single mode, the time evolution is often best visualized in the phase space \cite{S01}. First, we can use the expectation values of the Hermitian quadrature operators
\begin{equation}\label{XY}
X=\frac{a^\dagger+a}{2}, \ \ Y=i \frac{a^\dagger-a}{2}
\end{equation}
which, for localized states, coincides well with the motion of the center of the wave packet. (E.g., for a coherent state $\vert \alpha\rangle, $ we can easily see that the expectation value of $X$ $(Y)$ is the real (imaginary) part of $\alpha$.) However, in the general case these expectation values do not contain the complete quantum mechanical information that is available in a (reduced) density operator, $\rho.$ On the contrary, the Wigner function $W(\alpha)$ is in a one-to-one correspondence with $\rho.$ Using the photon number eigenstate expansion, $W$ is conveniently calculated as the expectation value of the Wigner operator \cite{BC95,Czirjak_1996}:
\begin{equation}\label{Wop}
\hat{W}(\alpha)=\frac{1}{\hbar \pi} D(\alpha)P D^\dagger(\alpha),
\end{equation}
where $D(\alpha)=\exp(\alpha a^\dagger - \alpha^* a)$ is the displacement operator, and $P$ denotes the parity. The function $W(\alpha)$ is obtained by utilizing the identities $PD(\alpha)=D(-\alpha)P$ and $P\vert n\rangle=(-1)^n \vert n\rangle$ together with the fact that $\langle n'| D(\alpha) |n\rangle$ can be expressed using an associated Laguerre function $L_n^{n-n'}(|\alpha|^2)$ \cite{WM94}.

\bigskip

Fig.~\ref{Wshort} summarizes the initial stage of the time evolution in phase space. (Note that in the context of parametric amplification \cite{MG67I,MG67II}, a similar figure with $P$ functions appeared already in Ref.~\cite{MG67I}.) In order to maintain the clarity of the figure, we set $\alpha_0=\sqrt{500}.$ For the third harmonic, the curve $(\langle X(t)\rangle,\langle Y(t)\rangle)$ is a spiral which is well described by the real and imaginary parts of the complex number $-it\beta(t)=-it\chi_3 \alpha_0^3 e^{-i 3 \omega t}.$ This supports the approximation that at this stage of the time evolution, the state of the HH modes are essentially the coherent states given by Eq.~(\ref{cohbeta}). The corresponding Wigner function is also very close to the Gaussian that corresponds to $\vert -i t\beta(t)\rangle$. This holds in spite of the fact that the exciting mode is not exactly the coherent state $\vert \alpha_0 e^{-i \omega t} \rangle,$ which is a direct consequence of losing photons that are upconverted to the HH mode.

However, the Wigner function that can be calculated using $\rho^{(0)}$ is still very close to a Gaussian. That is, in this stage of the time evolution the state of the exciting mode can be described well by a coherent state with an index the magnitude of which is decreasing.

\subsection{The complete dynamics}

\begin{figure}
\includegraphics[width=8.4cm]{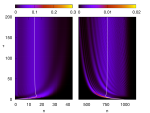}
\caption{Left panel: the photon number distribution $\Pi_n^{(15)}$ for the 15${}^{\mathrm{th}}$ HH mode as a function of time. Right panel:  $\Pi_n^{(0)}$  as a function of time. The solid white lines show the photon number expectation values (which are also the means of the respective distributions) in both panels. Parameters: $\alpha_0=\sqrt{1000},$ $p=20.0.$\label{longdistr}}
\end{figure}

Considering the efficiency and typical duration of the HHG process, the relative number of exciting mode photons  that are transferred to one of the HH modes is low, i.e., the photon number expectation value for the exciting mode is almost constant during the process. However, in order to complete the physical picture, it is instructive to investigate the time evolution on a considerably longer time scale.

Fig.~\ref{longdistr} shows the photon number distribution of the exciting mode and a single (the 15${}^{\mathrm{th}}$) HH mode as a function of time. As we can see, the initially narrow (Poissonian) distribution first gets structured, multiple peaks appear, then these peaks become less pronounced, eventually approaching a wide distribution. Correspondingly, the expectation value of the photon number operators (solid white line in Fig.~\ref{longdistr}) converges to constant values. Note that the initial quadratic increase of $\langle a_{15}^\dagger a_{15} \rangle$ is hardly visible in this figure, and similarly $\langle A^\dagger A \rangle$ decreases approximately linearly till its minimum, and then a slow convergence towards a constant value starts. Clearly, since the expectation value of $N$ given by Eq.~(\ref{Nop}) is a constant of motion, $\langle a_{15}^\dagger a_{15} \rangle$ is a (scaled) "mirror image" of $\langle A^\dagger A \rangle.$ The positions and actual values of the extrema depends on which HH mode is investigated.

{Explaining all the details of the complex dynamics shown in Fig.~\ref{longdistr} requires a thorough analysis. Considering the qualitative behavior, the time evolution can be divided into two parts. The initial stage (till $\tau\approx 10$) is quite intuitive, the photon distribution of the states remain localized (although super-Poissonian) and the dominant effect is the upconversion of exciting mode photons. However, since the number of these photons is finite, this process cannot last forever, and on the long timescale the flow of energy between the modes will be bidirectional. Thus the time evolution is analogous to the physical background of the "collapse and revival" phenomena that are often investigated in the context of a two-level atom that exchanges energy with an exciting field which is initially in a coherent state \cite{WM94}. In view of this, the fringes that can be seen in Fig.~\ref{longdistr} correspond to the phenomenon of "collapse", when the eigenstates of the complete system gradually lose the relative phases that initially led to localized photon number distributions. Note that the result of this dephasing can be seen in Fig.~\ref{finalW} as well. Our numerical calculations also show that the discrete frequencies that correspond to the time evolution of the eigenstates later can rephase again (not shown in the picture).  A similar "revival" process was shown to appear during parametric down-conversion in Ref.~\cite{DJ92}.}

\begin{figure}
\includegraphics[width=8.4cm]{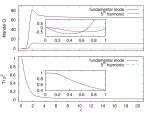}
\caption{Top panel: The Mandel parameter (\ref{Mandel}) as a function of time for the two modes indicated in the legend. Bottom panel: $\mathrm{Tr}\rho^2$ for the same two modes. Parameters: $\alpha_0=\sqrt{1500},$ $p=10.0.$ \label{long}}
\end{figure}

\bigskip

The difference of a photon number distribution from the Poissonian statistics is conveniently quantified using the Mandel parameter:
\begin{equation}\label{Mandel}
Q_n=\frac{\left\langle\left( a_n^\dagger a_n\right)^2\right\rangle - \left\langle a_n^\dagger a_n \right\rangle^2}{\left\langle a_n^\dagger a_n\right\rangle} -1.
\end{equation}
For a coherent state, $Q=0,$ while $Q>0$ ($Q<0$) corresponds to super-Poissonian (sub-Poissonian) distributions. The upper panel of Fig.~\ref{long} shows a typical example with the fundamental and the fifth HH mode. As we can see in the inset, at the initial stage of the time evolution $Q<0$ for both modes (indications to this effect can be seen already in Fig.~\ref{Ndistr}), but later on the variance of the photon numbers increases (in accord with Fig.~\ref{longdistr}), resulting in large positive values for $Q.$

Note that both Fig.~\ref{long} and Fig.~\ref{longdistr} suggest that the departure from the initial coherent state is faster for the fundamental mode than for the HH modes. This can be proven by a formal series expansion (see the Appendix), showing that for short times, $Q$ is a quadratic function of time for the fundamental mode, while its first nontrivial order is only $\tau^4$ for the HH modes. This behavior can be used for further analytic approximations.

\bigskip
{
Finally, we discuss quantum entaglement between the field modes.
Entanglement in strong-field and attosecond physics has obtained reviving attention in the past years \cite{Tzallas_2023}, including
also electon-ion \cite{Czirjak_2013,Majorosi_2017} and electron-electron entanglement \cite{Lewenstein_2022}.
The entanglement between our field modes is weak at the initial stage of
the time evolution, since the state of the system is appropriately
estimated by the tensor product of coherent
states.} This is visualized by the lower panel of Fig.~\ref{long}, in which $\mathrm{Tr}\left(\rho^{(0)}\right)^2$ and $\mathrm{Tr}\left(\rho^{(5)}\right)^2$ are plotted as a function of time. (Note that since in this case there are only two modes, we should have $\mathrm{Tr}\left(\rho^{(0)}\right)^2=\mathrm{Tr}\left(\rho^{(5)}\right)^2,$ which, as Fig.~\ref{long} shows, is true also in our numerical calculation.) As we can see, at the beginning of the time evolution $\mathrm{Tr}\left(\rho^{(0)}\right)^2=\mathrm{Tr}\left(\rho^{(5)}\right)^2$ are close to unity, indicating a tensor product state. However, later this quantity decreases considerably, i.e., a strong entanglement builds up in the long time limit.

\begin{figure}
\includegraphics[width=8.4cm]{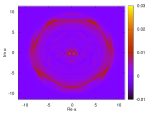}
\caption{The Wigner function corresponding to the 15${}^{\mathrm{th}}$ HH mode at the end of the time evolution ($\tau=200$) shown in Fig.~\ref{longdistr}. Note that the (radial) fwhm of a Gaussian Wigner function corresponding to a coherent state is $\sqrt{2\ln 2},$ see Fig.~\ref{Wshort}. The parameters are the same as for Fig.~\ref{longdistr}. \label{finalW}}
\end{figure}

\bigskip

The broadening of the photon number distribution have effects in the phase space as well. Fig.~\ref{finalW} shows the Wigner function corresponding to the 15${}^{\mathrm{th}}$ HH mode in the long time limit. As we can see, this is not a localized wave packet, $W(\alpha)$ rather has a circular symmetry. This indicates the loss of phase information during the time evolution (resulting in quadrature expectation values close to zero, not shown in the figure). Although -- according to Fig.~\ref{long} -- $\rho^{(15)}$ corresponding to Fig.~\ref{finalW} is not a pure state, weak negative fringes around $\alpha=0$ still indicate quantum mechanical interference and a nonclassical state.

\bigskip
That is, the time evolution of the system can be summarized as follows: At the initial stage the state of the system is close to a tensor product of coherent states. The magnitude of the labels for the HH mode coherent states are increasing linearly, while for the fundamental mode it is decreasing such that the "weighted photon number" [E.q.~(\ref{Nop})] is a constant of motion. When the photon number expectation value of the exciting mode becomes considerable lower than its initial value, this picture qualitatively changes, mode-mode entanglement builds up and all photon number distributions get broader. Also in the phase space, the Wigner functions become less localized. Or in terms of the eigenstates of the system:  their initial superpositions lead to localized phase space distributions, but later on, as the phases of the eigenfunctions evolve, the distributions broaden. Since the problem is finite dimensional with discrete spectrum, the eigenstates can rephase leading to partial revivals.

\section{Summary}
In the current paper we introduced a model for high-order harmonic generation in which both the exciting and the harmonic modes are treated as quantized fields. We adapted the Hamiltonian used for describing the process of parametric down-conversion to this problem by considering several HH modes. For an initial state which is the product of a coherent state (excitation) and vacuum states (HH modes), it was shown analytically that at the initial stage of the time evolution all fields are in coherent states. The parameters of these coherent states were explicitly given and verified by numerical calculations. For longer times, numerical results indicated a deviation from this behavior. The photon number distributions get broader, and -- as it was shown by Wigner functions -- localization in the phase space becomes less pronounced.

\section*{Acknowledgments}
Partial support by the ELI-ALPS project is acknowledged.
The ELI-ALPS project (GINOP-2.3.6-15-2015-00001) is supported
by the European Union and co-financed by the European Regional Development Fund.

\'A.~G. acknowledges support from the project no. TKP2021-NVA-04 of the Ministry of Innovation and Technology of Hungary.

The work of B.G. Pusztai was supported by project TKP2021-NVA-09. Project no. TKP2021-NVA-09 has been implemented with the support provided by the Ministry of Innovation and Technology of Hungary from the National Research, Development and Innovation Fund, financed under the TKP2021-NVA funding scheme.

{We thank Zsolt Div\'eki for the experimental data corresponding to Fig.~7.}

\appendix*
\subsection{On the photon number statistics}
In this appendix we shall study the time evolution of the photon number
statistics for each mode. From this perspective it is crucial that the
weighted photon number operator \eqref{Nop} and the interaction term
\eqref{H_I} commute, and so the evolution operator associated with the
Hamiltonian \eqref{Ham} factorizes as
\begin{equation}\label{APP__U(t)}
    U(t) = e^{-i \omega t N} e^{-\frac{i}{\hbar} t H_I}.
\end{equation}
To proceed further, let $P_n$ denote the parity operator of the HH mode
$n$. Upon introducing their product,
\begin{equation}\label{APP__cP}
    \mathcal{P} = \prod_n P_n,
\end{equation}
a moment of reflection reveals that
\begin{equation}
    N \mathcal{P} = \mathcal{P} N
    \quad\text{and}\quad
    H_I \mathcal{P} = -\mathcal{P} H_I,
\end{equation}
from where we obtain at once that
\begin{equation}\label{APP__U(t)_cP}
    U(t) \mathcal{P} = \mathcal{P} e^{-2 i \omega t N} U(-t).
\end{equation}
Turning to the photon number operators, clearly
\begin{equation}
    [A^\dagger A, \mathcal{P}] = 0
    \quad\text{and}\quad
    [a_n^\dagger a_n, \mathcal{P}] = 0.
\end{equation}
Consequently, the relationship \eqref{APP__U(t)_cP} entails that for all
non-negative integer $r$ we have
\begin{equation}\label{APP__number_op_powers_and_U(t)}
    \mathcal{P}^\dagger U(t)^\dagger (A^\dagger A)^r U(t) \mathcal{P}
    = U(-t)^\dagger (A^\dagger A)^r U(-t),
\end{equation}
and an analogous formula holds for each HH mode, too, simply by replacing
letter $A$ with $a_n$.

Now, our investigation hinges on the fact that the initial state
\eqref{initial} is compatible with $\mathcal{P}$ in the sense that
\begin{equation}
    \mathcal{P} \ket{\Psi} = \ket{\Psi}.
\end{equation}
Therefore, by exploiting \eqref{APP__number_op_powers_and_U(t)}, along the
time evolution of $\ket{\Psi}$ for the $r$th moment of the photon number
statistics in the fundamental mode we get
\begin{equation}
\begin{aligned}
    \braket{\Psi | U(t)^\dagger (A^\dagger A)^r U(t) | \Psi}
    & =
    \braket{\Psi | \mathcal{P}^\dagger U(t)^\dagger (A^\dagger A)^r U(t) \mathcal{P}| \Psi}
    \\
    & =
    \braket{\Psi | U(-t)^\dagger (A^\dagger A)^r U(-t) | \Psi}.
\end{aligned}
\end{equation}
That is, the upshot of our symmetry argument is that each moment of the
photon numbers statistics in the fundamental mode is an \emph{even} function
of time $t$. Of course, the same argument applies to the HH modes as well.
Giving a glance at \eqref{Mandel}, it readily follows that the time
dependence of the Mandel $Q$ parameter is also even. In particular, around
$t = 0$, the power series expansions of the moments and the Mandel
parameters contain only terms of even degree.

Keeping in mind the above observations, in the rest of the appendix we
examine the short time behavior of the Mandel parameter for each mode. We
perform the calculations in the Heisenberg picture, but for simplicity we
omit the usual $H$ subscript from the time dependent operators. With this
proviso, from \eqref{Ham} it is plain that the time evolution of the
annihilation operators is governed by the following system of differential
equations:
\begin{align}
    & \dot{A}(t)
    = - i \omega A(t)
        - i \sum_n n \chi_n (A(t)^\dagger)^{n - 1} a_n(t),
    \label{APP__dot_A}
    \\
    & \dot{a}_n(t) = - i n \omega a_n(t) - i \chi_n A(t)^n.
    \label{APP__dot_a_n}
\end{align}
Note that, in terms of the auxiliary operators
\begin{equation}\label{APP__B_and_b_n}
    B(t) = e^{i \omega t} A(t)
    \quad\text{and}\quad
    b_n(t) = e^{i n \omega t} a_n(t),
\end{equation}
the dynamics takes the somewhat simpler form
\begin{align}
    & \dot{B}(t)
    = - i \sum_n n \chi_n (B(t)^\dagger)^{n - 1} b_n(t),
    \label{APP__B_eq}
    \\
    & \dot{b}_n(t) = - i \chi_n B(t)^n.
    \label{APP__b_n_eq}
\end{align}
Furthermore, for the first two moments of the photon number statistics
in the fundamental mode we can write
\begin{align}
    & \braket{A(t)^\dagger A(t)}
    = \braket{B(t)^\dagger B(t)},
    \label{APP__A_B_1st_moment} \\
    & \braket{(A(t)^\dagger A(t))^2}
    = \braket{(B(t)^2)^\dagger B(t)^2}
     + \braket{B(t)^\dagger B(t)}.
     \label{APP__A_B_2nd_moment}
\end{align}
Of course, analogous formulae hold for the HH modes as well.

Since the system displayed in \eqref{APP__B_eq} and \eqref{APP__b_n_eq}
is highly non-linear, we do not expect that its solution is expressible
in closed form. Nevertheless, in order to capture the short time behavior
of the photon number statistics, it is enough to work out the first few
coefficients of the power series
\begin{equation}\label{APP__power_series}
    B(t) = \sum_{k = 0}^\infty t^k B^{(k)}
    \quad\text{and}\quad
    b_n(t) = \sum_{k = 0}^\infty t^k b_n^{(k)}.
\end{equation}
Starting with the study of the fundamental mode, elementary calculations
lead to the formulae
\begin{equation}
    B^{(0)} = A
    \quad\text{and}\quad
    B^{(1)} = -i \sum_n n \chi_n (A^{n - 1})^\dagger a_n,
\end{equation}
whereas for the second order coefficient we obtain
\begin{equation}
\begin{aligned}
    B^{(2)} =
        & -\frac{1}{2} \sum_n n \chi_n^2 (A^{n - 1})^\dagger A^n \\
        & + \frac{1}{2}
            \sum_{n, n'}
            \sum_{s = 0}^{n - 2}
                n n' \chi_n \chi_{n'}
                (A^s)^\dagger A^{n' - 1} (A^{n - 2 - s})^\dagger
                a_{n'}^\dagger a_n.
\end{aligned}
\end{equation}
Note that on the right hand sides of the above equations the operators
$A$ and $a_n$ refer to the annihilation operators in the Schrödinger
picture. Since for their action on the initial state \eqref{initial}
we have
\begin{equation}
    A \ket{\Psi} = \alpha_0 \ket{\Psi}
    \quad \text{and} \quad
    a_n \ket{\Psi} = 0,
\end{equation}
for the time evolution of the expectation value
\eqref{APP__A_B_1st_moment} of the photon numbers in the fundamental mode
we arrive at
\begin{equation}
    \braket{A(t)^\dagger A(t)} =
    \vert \alpha_0 \vert^2
        - t^2 \sum_n n \chi_n^2 \vert \alpha_0 \vert^{2 n}
        + \mathcal{O}(t^4).
\end{equation}
Furthermore, utilizing \eqref{APP__A_B_2nd_moment}, for the second
moment we find
\begin{equation}
\begin{aligned}
   \braket{(A(t)^\dagger A(t))^2} =
    & \vert \alpha_0 \vert^4 + \vert \alpha_0 \vert^2 \\
    & - t^2 \sum_n \chi_n^2 \vert \alpha_0 \vert^{2 n}
            \left( n^2 + 2 n \vert \alpha_0 \vert^2 \right) \\
    & + \mathcal{O}(t^4).
\end{aligned}
\end{equation}
Now, by plugging the above formulae into \eqref{Mandel}, for the Mandel
parameter in the fundamental mode we end up with
\begin{equation}
    Q_0(t) =
    - t^2 \sum_n (n^2 - n) \chi_n^2 \vert \alpha_0 \vert^{2 n - 2}
    + \mathcal{O}(t^4).
\end{equation}

We complete this appendix by examining the photon number statistics in
the HH modes, too. It turns out that, contrary to the
fundamental mode, the quadratic approximation in the power series
\eqref{APP__power_series} is not sufficient to derive the leading order
behavior of the Mandel parameter. Nevertheless, by going one stage up,
for the HH mode $n$ the cubic approximation of the power series leads
to the estimation
\begin{equation}
    Q_n(t) = \mathcal{O}(t^4).
\end{equation}
However, the full control over the leading order behavior of $Q_n(t)$ does
require even the fourth order terms in \eqref{APP__power_series}, thus
the calculations are feasible only in the simplest special case of the
second harmonic generation. Indeed, under the assumption that there is
only a single HH mode characterized by $n = 2$, one can show that
\begin{equation}
    Q_2(t) =
    -\frac{4}{3} t^4 \chi_2^4 \vert \alpha_0 \vert^4
    + \mathcal{O}(t^6).
\end{equation}

\begin{figure}
\includegraphics[width=8.4cm]{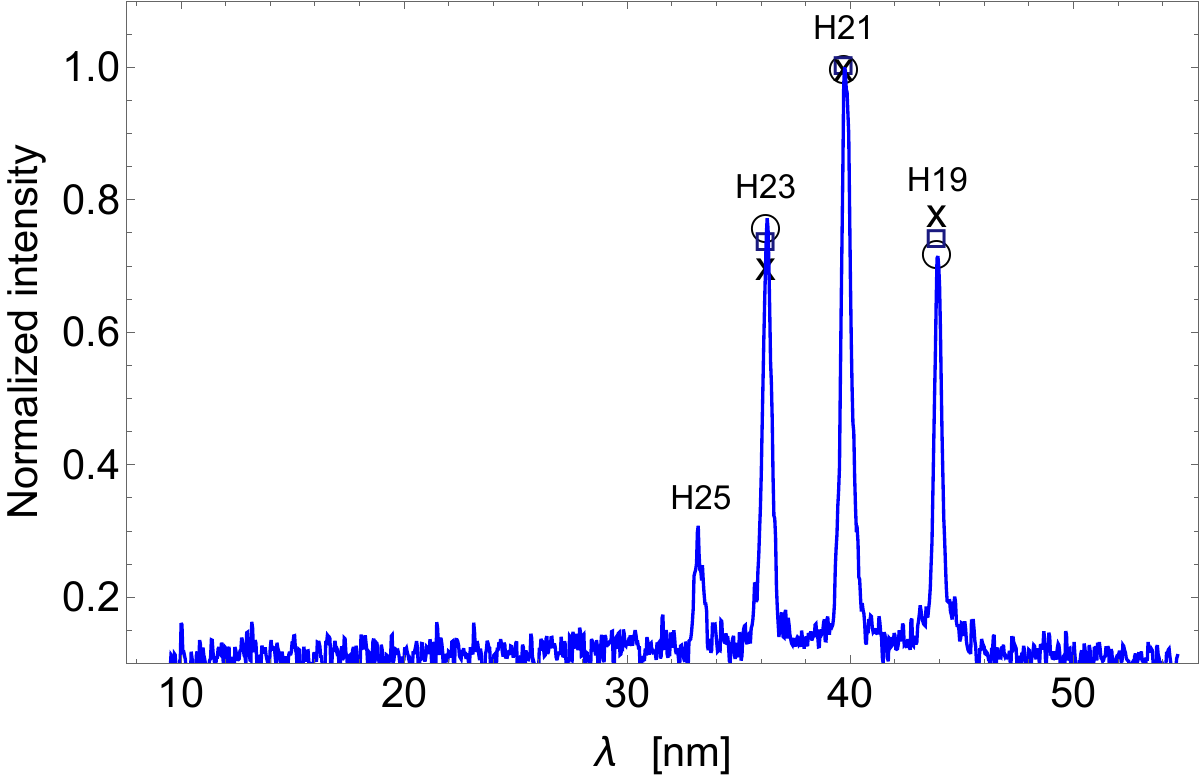}
\caption{{A part of the HHG spectrum of the argon gas. (Excitation: 17.2 mJ energy at $\lambda=834$ nm, duration: 28.1 fs intensity fwhm.) The data are normalized so that the height of the 21$^{\mathrm {st}}$ HH peak is unity.  The symbols correspond to calculated expectation values of the "weighted photon number operators"  $n \langle a_{n}^\dagger a_{n}\rangle$ at time instants when the photon number expectation value in the exciting mode becomes 98\% (circles), 95\% (squares) and 90\% (crosses) of its initial value (which was 1000).  For more details, see the main text. \label{expspectr}}}
\end{figure}

{
\subsection{Modeling experimental results}
Finally, let us consider a specific experimental example. Fig.~\ref{expspectr} shows a part of the spectrum of argon gas that was excited by a pulse with 17.2 mJ energy at $\lambda=834$ nm (duration: 28.1 fs intensity fwhm) using the SYLOS system of the ELI-ALPS \cite{Appi23}. The heights of the HH peaks are clearly not equal in this case, but we can choose appropriate parameters $\chi_{19}, \chi_{21}, \chi_{23}$  and $\chi_{25}$ to reproduce these peaks. Clearly, the parameter $p$ given by Eq.~(\ref{Pparam}) cannot be constant in this case. Instead (according to the calculation in Sec.~\ref{Paramsec}) we choose e.g.:
\begin{equation}
  \frac{\chi_{21}}{\chi_{n}} = \frac{\sqrt{n}|\alpha_0|^{n}}{\sqrt{h_n \times 21} |\alpha_0|^{21}},
  \label{chiexp}
\end{equation}
where the numerical factors $h_n$ are the relative heights of the corresponding peaks in Fig.~\ref{expspectr} (as compared to the 21$^{\mathrm {st}}$ one.) Focusing on $n=19$ and 23 ($h_{19}=0.71, h_{23}=0.77$), we performed numerical calculations with $\alpha_0=\sqrt{1000}.$ We used Eq.~(\ref{chiexp}), and -- as before -- we considered two harmonic modes at a time, i.e., the pairs (21,19) and (21,23) were investigated. In both cases we found that the behavior of $21 \langle a_{21}^\dagger a_{21}\rangle (\tau)$ and $n \langle a_{n}^\dagger a_{n}\rangle (\tau)$ ($n=19, 23$) is quadratic in $\tau$ (c.f.~Fig \ref{Nexpini}), and their ratio is close to $h_n$ at the beginning of the time evolution. More precisely, e.g., at the time instant when the photon number expectation value of the exciting mode dropped to 90\% of its initial value, the relative difference between  $21 \langle a_{21}^\dagger a_{21}\rangle (\tau) /[n \langle a_{n}^\dagger a_{n}\rangle (\tau)]$ and $h_n$ was below 10\%.}

{These calculations mean a direct connection between our model and experimental results, and demonstrate the flexibility of the approach introduced earlier in this paper.}


\begin{thebibliography}{0}%
\makeatletter
\providecommand \@ifxundefined [1]{%
 \@ifx{#1\undefined}
}%
\providecommand \@ifnum [1]{%
 \ifnum #1\expandafter \@firstoftwo
 \else \expandafter \@secondoftwo
 \fi
}%
\providecommand \@ifx [1]{%
 \ifx #1\expandafter \@firstoftwo
 \else \expandafter \@secondoftwo
 \fi
}%
\providecommand \natexlab [1]{#1}%
\providecommand \enquote  [1]{``#1''}%
\providecommand \bibnamefont  [1]{#1}%
\providecommand \bibfnamefont [1]{#1}%
\providecommand \citenamefont [1]{#1}%
\providecommand \href@noop [0]{\@secondoftwo}%
\providecommand \href [0]{\begingroup \@sanitize@url \@href}%
\providecommand \@href[1]{\@@startlink{#1}\@@href}%
\providecommand \@@href[1]{\endgroup#1\@@endlink}%
\providecommand \@sanitize@url [0]{\catcode `\\12\catcode `\$12\catcode
  `\&12\catcode `\#12\catcode `\^12\catcode `\_12\catcode `\%12\relax}%
\providecommand \@@startlink[1]{}%
\providecommand \@@endlink[0]{}%
\providecommand \url  [0]{\begingroup\@sanitize@url \@url }%
\providecommand \@url [1]{\endgroup\@href {#1}{\urlprefix }}%
\providecommand \urlprefix  [0]{URL }%
\providecommand \Eprint [0]{\href }%
\providecommand \doibase [0]{https://doi.org/}%
\providecommand \selectlanguage [0]{\@gobble}%
\providecommand \bibinfo  [0]{\@secondoftwo}%
\providecommand \bibfield  [0]{\@secondoftwo}%
\providecommand \translation [1]{[#1]}%
\providecommand \BibitemOpen [0]{}%
\providecommand \bibitemStop [0]{}%
\providecommand \bibitemNoStop [0]{.\EOS\space}%
\providecommand \EOS [0]{\spacefactor3000\relax}%
\providecommand \BibitemShut  [1]{\csname bibitem#1\endcsname}%
\let\auto@bib@innerbib\@empty
\end{thebibliography}%


\begin{thebibliography}{45}%
\makeatletter
\providecommand \@ifxundefined [1]{%
 \@ifx{#1\undefined}
}%
\providecommand \@ifnum [1]{%
 \ifnum #1\expandafter \@firstoftwo
 \else \expandafter \@secondoftwo
 \fi
}%
\providecommand \@ifx [1]{%
 \ifx #1\expandafter \@firstoftwo
 \else \expandafter \@secondoftwo
 \fi
}%
\providecommand \natexlab [1]{#1}%
\providecommand \enquote  [1]{``#1''}%
\providecommand \bibnamefont  [1]{#1}%
\providecommand \bibfnamefont [1]{#1}%
\providecommand \citenamefont [1]{#1}%
\providecommand \href@noop [0]{\@secondoftwo}%
\providecommand \href [0]{\begingroup \@sanitize@url \@href}%
\providecommand \@href[1]{\@@startlink{#1}\@@href}%
\providecommand \@@href[1]{\endgroup#1\@@endlink}%
\providecommand \@sanitize@url [0]{\catcode `\\12\catcode `\$12\catcode
  `\&12\catcode `\#12\catcode `\^12\catcode `\_12\catcode `\%12\relax}%
\providecommand \@@startlink[1]{}%
\providecommand \@@endlink[0]{}%
\providecommand \url  [0]{\begingroup\@sanitize@url \@url }%
\providecommand \@url [1]{\endgroup\@href {#1}{\urlprefix }}%
\providecommand \urlprefix  [0]{URL }%
\providecommand \Eprint [0]{\href }%
\providecommand \doibase [0]{https://doi.org/}%
\providecommand \selectlanguage [0]{\@gobble}%
\providecommand \bibinfo  [0]{\@secondoftwo}%
\providecommand \bibfield  [0]{\@secondoftwo}%
\providecommand \translation [1]{[#1]}%
\providecommand \BibitemOpen [0]{}%
\providecommand \bibitemStop [0]{}%
\providecommand \bibitemNoStop [0]{.\EOS\space}%
\providecommand \EOS [0]{\spacefactor3000\relax}%
\providecommand \BibitemShut  [1]{\csname bibitem#1\endcsname}%
\let\auto@bib@innerbib\@empty
\bibitem [{\citenamefont {McPherson}\ \emph {et~al.}(1987)\citenamefont
  {McPherson}, \citenamefont {Gibson}, \citenamefont {Jara}, \citenamefont
  {Johann}, \citenamefont {Luk}, \citenamefont {McIntyre}, \citenamefont
  {Boyer},\ and\ \citenamefont {Rhodes}}]{McPherson87}%
  \BibitemOpen
  \bibfield  {author} {\bibinfo {author} {\bibfnamefont {A.}~\bibnamefont
  {McPherson}}, \bibinfo {author} {\bibfnamefont {G.}~\bibnamefont {Gibson}},
  \bibinfo {author} {\bibfnamefont {H.}~\bibnamefont {Jara}}, \bibinfo {author}
  {\bibfnamefont {U.}~\bibnamefont {Johann}}, \bibinfo {author} {\bibfnamefont
  {T.~S.}\ \bibnamefont {Luk}}, \bibinfo {author} {\bibfnamefont {I.~A.}\
  \bibnamefont {McIntyre}}, \bibinfo {author} {\bibfnamefont {K.}~\bibnamefont
  {Boyer}},\ and\ \bibinfo {author} {\bibfnamefont {C.~K.}\ \bibnamefont
  {Rhodes}},\ }\bibfield  {title} {\bibinfo {title} {Studies of multiphoton
  production of vacuum-ultraviolet radiation in the rare gases},\ }\href@noop
  {} {\bibfield  {journal} {\bibinfo  {journal} {J. Opt. Soc. Am. B}\ }\textbf
  {\bibinfo {volume} {4}},\ \bibinfo {pages} {595} (\bibinfo {year}
  {1987})}\BibitemShut {NoStop}%
\bibitem [{\citenamefont {Ferray}\ \emph {et~al.}(1988)\citenamefont {Ferray},
  \citenamefont {L'Huillier}, \citenamefont {Li}, \citenamefont {Lompre},
  \citenamefont {Mainfray},\ and\ \citenamefont {Manus}}]{Ferray88}%
  \BibitemOpen
  \bibfield  {author} {\bibinfo {author} {\bibfnamefont {M.}~\bibnamefont
  {Ferray}}, \bibinfo {author} {\bibfnamefont {A.}~\bibnamefont {L'Huillier}},
  \bibinfo {author} {\bibfnamefont {X.~F.}\ \bibnamefont {Li}}, \bibinfo
  {author} {\bibfnamefont {L.~A.}\ \bibnamefont {Lompre}}, \bibinfo {author}
  {\bibfnamefont {G.}~\bibnamefont {Mainfray}},\ and\ \bibinfo {author}
  {\bibfnamefont {C.}~\bibnamefont {Manus}},\ }\bibfield  {title} {\bibinfo
  {title} {Multiple-harmonic conversion of 1064 nm radiation in rare gases},\
  }\href@noop {} {\bibfield  {journal} {\bibinfo  {journal} {J. Phys. B: At.
  Mol. Phys.}\ }\textbf {\bibinfo {volume} {21}},\ \bibinfo {pages} {L31}
  (\bibinfo {year} {1988})}\BibitemShut {NoStop}%
\bibitem [{\citenamefont {Farkas}\ and\ \citenamefont
  {T\'{o}th}(1992)}]{FT1992}%
  \BibitemOpen
  \bibfield  {author} {\bibinfo {author} {\bibfnamefont {G.}~\bibnamefont
  {Farkas}}\ and\ \bibinfo {author} {\bibfnamefont {C.}~\bibnamefont
  {T\'{o}th}},\ }\bibfield  {title} {\bibinfo {title} {Proposal for attosecond
  light pulse generation using laser induced multiple-harmonic conversion
  processes in rare gases},\ }\href@noop {} {\bibfield  {journal} {\bibinfo
  {journal} {Phys. Lett. A}\ }\textbf {\bibinfo {volume} {168}},\ \bibinfo
  {pages} {447} (\bibinfo {year} {1992})}\BibitemShut {NoStop}%
\bibitem [{\citenamefont {Qu\'er\'e}\ \emph {et~al.}(2006)\citenamefont
  {Qu\'er\'e}, \citenamefont {Thaury}, \citenamefont {Monot}, \citenamefont
  {Dobosz}, \citenamefont {Martin}, \citenamefont {Geindre},\ and\
  \citenamefont {Audebert}}]{QTMDMGA2006}%
  \BibitemOpen
  \bibfield  {author} {\bibinfo {author} {\bibfnamefont {F.}~\bibnamefont
  {Qu\'er\'e}}, \bibinfo {author} {\bibfnamefont {C.}~\bibnamefont {Thaury}},
  \bibinfo {author} {\bibfnamefont {P.}~\bibnamefont {Monot}}, \bibinfo
  {author} {\bibfnamefont {S.}~\bibnamefont {Dobosz}}, \bibinfo {author}
  {\bibfnamefont {P.}~\bibnamefont {Martin}}, \bibinfo {author} {\bibfnamefont
  {J.-P.}\ \bibnamefont {Geindre}},\ and\ \bibinfo {author} {\bibfnamefont
  {P.}~\bibnamefont {Audebert}},\ }\bibfield  {title} {\bibinfo {title}
  {Coherent wake emission of high-order harmonics from overdense plasmas},\
  }\href@noop {} {\bibfield  {journal} {\bibinfo  {journal} {Phys. Rev. Lett.}\
  }\textbf {\bibinfo {volume} {96}},\ \bibinfo {pages} {125004} (\bibinfo
  {year} {2006})}\BibitemShut {NoStop}%
\bibitem [{\citenamefont {Vincenti}\ \emph {et~al.}(2014)\citenamefont
  {Vincenti}, \citenamefont {Monchoc{\'e}}, \citenamefont {Kahaly},
  \citenamefont {Bonnaud}, \citenamefont {Martin},\ and\ \citenamefont
  {Qu{\'e}r{\'e}}}]{Vincenti2014}%
  \BibitemOpen
  \bibfield  {author} {\bibinfo {author} {\bibfnamefont {H.}~\bibnamefont
  {Vincenti}}, \bibinfo {author} {\bibfnamefont {S.}~\bibnamefont
  {Monchoc{\'e}}}, \bibinfo {author} {\bibfnamefont {S.}~\bibnamefont
  {Kahaly}}, \bibinfo {author} {\bibfnamefont {G.}~\bibnamefont {Bonnaud}},
  \bibinfo {author} {\bibfnamefont {P.}~\bibnamefont {Martin}},\ and\ \bibinfo
  {author} {\bibfnamefont {F.}~\bibnamefont {Qu{\'e}r{\'e}}},\ }\bibfield
  {title} {\bibinfo {title} {Optical properties of relativistic plasma
  mirrors},\ }\href@noop {} {\bibfield  {journal} {\bibinfo  {journal} {Nat.
  Comm.}\ }\textbf {\bibinfo {volume} {5}},\ \bibinfo {pages} {3403} (\bibinfo
  {year} {2014})}\BibitemShut {NoStop}%
\bibitem [{\citenamefont {Ghimire}\ \emph {et~al.}(2011)\citenamefont
  {Ghimire}, \citenamefont {DiChiara}, \citenamefont {Sistrunk}, \citenamefont
  {Agostini}, \citenamefont {DiMauro},\ and\ \citenamefont
  {Reis}}]{GDiCSADiMR11}%
  \BibitemOpen
  \bibfield  {author} {\bibinfo {author} {\bibfnamefont {S.}~\bibnamefont
  {Ghimire}}, \bibinfo {author} {\bibfnamefont {A.~D.}\ \bibnamefont
  {DiChiara}}, \bibinfo {author} {\bibfnamefont {E.}~\bibnamefont {Sistrunk}},
  \bibinfo {author} {\bibfnamefont {P.}~\bibnamefont {Agostini}}, \bibinfo
  {author} {\bibfnamefont {L.~F.}\ \bibnamefont {DiMauro}},\ and\ \bibinfo
  {author} {\bibfnamefont {D.~A.}\ \bibnamefont {Reis}},\ }\bibfield  {title}
  {\bibinfo {title} {Observation of high-order harmonic generation in a bulk
  crystal},\ }\href@noop {} {\bibfield  {journal} {\bibinfo  {journal} {Nat.
  Phys.}\ }\textbf {\bibinfo {volume} {7}},\ \bibinfo {pages} {138} (\bibinfo
  {year} {2011})}\BibitemShut {NoStop}%
\bibitem [{\citenamefont {Ghimire}\ and\ \citenamefont {Reis}(2019)}]{Ghim19}%
  \BibitemOpen
  \bibfield  {author} {\bibinfo {author} {\bibfnamefont {S.}~\bibnamefont
  {Ghimire}}\ and\ \bibinfo {author} {\bibfnamefont {D.~A.}\ \bibnamefont
  {Reis}},\ }\bibfield  {title} {\bibinfo {title} {High-harmonic generation
  from solids},\ }\href@noop {} {\bibfield  {journal} {\bibinfo  {journal}
  {Nat. Phys.}\ }\textbf {\bibinfo {volume} {15}},\ \bibinfo {pages} {10}
  (\bibinfo {year} {2019})}\BibitemShut {NoStop}%
\bibitem [{\citenamefont {Corkum}(1993)}]{Cor93}%
  \BibitemOpen
  \bibfield  {author} {\bibinfo {author} {\bibfnamefont {P.~B.}\ \bibnamefont
  {Corkum}},\ }\bibfield  {title} {\bibinfo {title} {Plasma perspective on
  strong field multiphoton ionization},\ }\href@noop {} {\bibfield  {journal}
  {\bibinfo  {journal} {Phys. Rev. Lett.}\ }\textbf {\bibinfo {volume} {71}},\
  \bibinfo {pages} {1994} (\bibinfo {year} {1993})}\BibitemShut {NoStop}%
\bibitem [{\citenamefont {Lewenstein}\ \emph {et~al.}(1994)\citenamefont
  {Lewenstein}, \citenamefont {Balcou}, \citenamefont {Ivanov}, \citenamefont
  {L'Huillier},\ and\ \citenamefont {Corkum}}]{Lew94}%
  \BibitemOpen
  \bibfield  {author} {\bibinfo {author} {\bibfnamefont {M.}~\bibnamefont
  {Lewenstein}}, \bibinfo {author} {\bibfnamefont {P.}~\bibnamefont {Balcou}},
  \bibinfo {author} {\bibfnamefont {M.~Y.}\ \bibnamefont {Ivanov}}, \bibinfo
  {author} {\bibfnamefont {A.}~\bibnamefont {L'Huillier}},\ and\ \bibinfo
  {author} {\bibfnamefont {P.~B.}\ \bibnamefont {Corkum}},\ }\bibfield  {title}
  {\bibinfo {title} {Theory of high-harmonic generation by low-frequency laser
  fields},\ }\href@noop {} {\bibfield  {journal} {\bibinfo  {journal} {Phys.
  Rev. A}\ }\textbf {\bibinfo {volume} {49}},\ \bibinfo {pages} {2117}
  (\bibinfo {year} {1994})}\BibitemShut {NoStop}%
\bibitem [{\citenamefont {Tsatrafyllis}\ \emph {et~al.}(2017)\citenamefont
  {Tsatrafyllis}, \citenamefont {Kominis}, \citenamefont {Gonoskov},\ and\
  \citenamefont {Tzallas}}]{Tsatrafyllis2017}%
  \BibitemOpen
  \bibfield  {author} {\bibinfo {author} {\bibfnamefont {N.}~\bibnamefont
  {Tsatrafyllis}}, \bibinfo {author} {\bibfnamefont {I.~K.}\ \bibnamefont
  {Kominis}}, \bibinfo {author} {\bibfnamefont {I.~A.}\ \bibnamefont
  {Gonoskov}},\ and\ \bibinfo {author} {\bibfnamefont {P.}~\bibnamefont
  {Tzallas}},\ }\bibfield  {title} {\bibinfo {title} {High-order harmonics
  measured by the photon statistics of the infrared driving-field exiting the
  atomic medium},\ }\href@noop {} {\bibfield  {journal} {\bibinfo  {journal}
  {Nat. Comm.}\ }\textbf {\bibinfo {volume} {8}} (\bibinfo {year}
  {2017})}\BibitemShut {NoStop}%
\bibitem [{\citenamefont {Tsatrafyllis}\ \emph {et~al.}(2019)\citenamefont
  {Tsatrafyllis}, \citenamefont {K\"{u}hn}, \citenamefont {Dumergue},
  \citenamefont {F\"{o}ldi}, \citenamefont {Kahaly}, \citenamefont {Cormier},
  \citenamefont {Gonoskov}, \citenamefont {Kiss}, \citenamefont {Varj\'{u}},
  \citenamefont {Varr\'{o}},\ and\ \citenamefont {Tzallas}}]{T19}%
  \BibitemOpen
  \bibfield  {author} {\bibinfo {author} {\bibfnamefont {N.}~\bibnamefont
  {Tsatrafyllis}}, \bibinfo {author} {\bibfnamefont {S.}~\bibnamefont
  {K\"{u}hn}}, \bibinfo {author} {\bibfnamefont {M.}~\bibnamefont {Dumergue}},
  \bibinfo {author} {\bibfnamefont {P.}~\bibnamefont {F\"{o}ldi}}, \bibinfo
  {author} {\bibfnamefont {S.}~\bibnamefont {Kahaly}}, \bibinfo {author}
  {\bibfnamefont {E.}~\bibnamefont {Cormier}}, \bibinfo {author} {\bibfnamefont
  {I.}~\bibnamefont {Gonoskov}}, \bibinfo {author} {\bibfnamefont
  {B.}~\bibnamefont {Kiss}}, \bibinfo {author} {\bibfnamefont {K.}~\bibnamefont
  {Varj\'{u}}}, \bibinfo {author} {\bibfnamefont {S.}~\bibnamefont
  {Varr\'{o}}},\ and\ \bibinfo {author} {\bibfnamefont {P.}~\bibnamefont
  {Tzallas}},\ }\bibfield  {title} {\bibinfo {title} {Quantum optical
  signatures in a strong laser pulse after interaction with semiconductors},\
  }\href@noop {} {\bibfield  {journal} {\bibinfo  {journal} {Phys. Rev. Lett.}\
  }\textbf {\bibinfo {volume} {122}},\ \bibinfo {pages} {193602} (\bibinfo
  {year} {2019})}\BibitemShut {NoStop}%
\bibitem [{\citenamefont {Bergou}\ and\ \citenamefont
  {Varr\'{o}}(1981{\natexlab{a}})}]{BV81b}%
  \BibitemOpen
  \bibfield  {author} {\bibinfo {author} {\bibfnamefont {J.}~\bibnamefont
  {Bergou}}\ and\ \bibinfo {author} {\bibfnamefont {S.}~\bibnamefont
  {Varr\'{o}}},\ }\bibfield  {title} {\bibinfo {title} {Nonlinear scattering
  processes in the presence of a quantised radiation field. II. Relativistic
  treatment},\ }\href@noop {} {\bibfield  {journal} {\bibinfo  {journal} {J.
  Phys. A: Math. Gen.}\ }\textbf {\bibinfo {volume} {14}},\ \bibinfo {pages}
  {2281} (\bibinfo {year} {1981}{\natexlab{a}})}\BibitemShut {NoStop}%
\bibitem [{\citenamefont {Bergou}\ and\ \citenamefont
  {Varr\'{o}}(1981{\natexlab{b}})}]{BV81a}%
  \BibitemOpen
  \bibfield  {author} {\bibinfo {author} {\bibfnamefont {J.}~\bibnamefont
  {Bergou}}\ and\ \bibinfo {author} {\bibfnamefont {S.}~\bibnamefont
  {Varr\'{o}}},\ }\bibfield  {title} {\bibinfo {title} {Nonlinear scattering
  processes in the presence of a quantised radiation field. I. Non-relativistic
  treatment},\ }\href@noop {} {\bibfield  {journal} {\bibinfo  {journal} {J.
  Phys. A: Math. Gen.}\ }\textbf {\bibinfo {volume} {14}},\ \bibinfo {pages}
  {1469} (\bibinfo {year} {1981}{\natexlab{b}})}\BibitemShut {NoStop}%
\bibitem [{\citenamefont {Gao}\ \emph {et~al.}(1998)\citenamefont {Gao},
  \citenamefont {Shen},\ and\ \citenamefont {Eden}}]{GSE98}%
  \BibitemOpen
  \bibfield  {author} {\bibinfo {author} {\bibfnamefont {J.}~\bibnamefont
  {Gao}}, \bibinfo {author} {\bibfnamefont {F.}~\bibnamefont {Shen}},\ and\
  \bibinfo {author} {\bibfnamefont {J.~G.}\ \bibnamefont {Eden}},\ }\bibfield
  {title} {\bibinfo {title} {Quantum electrodynamic treatment of harmonic
  generation in intense optical fields},\ }\href@noop {} {\bibfield  {journal}
  {\bibinfo  {journal} {Phys. Rev. Lett.}\ }\textbf {\bibinfo {volume} {81}},\
  \bibinfo {pages} {1833} (\bibinfo {year} {1998})}\BibitemShut {NoStop}%
\bibitem [{\citenamefont {Chen}\ \emph {et~al.}(2000)\citenamefont {Chen},
  \citenamefont {Chen},\ and\ \citenamefont {Liu}}]{CCL00}%
  \BibitemOpen
  \bibfield  {author} {\bibinfo {author} {\bibfnamefont {J.}~\bibnamefont
  {Chen}}, \bibinfo {author} {\bibfnamefont {S.~G.}\ \bibnamefont {Chen}},\
  and\ \bibinfo {author} {\bibfnamefont {J.}~\bibnamefont {Liu}},\ }\bibfield
  {title} {\bibinfo {title} {Comment on ``quantum electrodynamic treatment of
  harmonic generation in intense optical fields''},\ }\href@noop {} {\bibfield
  {journal} {\bibinfo  {journal} {Phys. Rev. Lett.}\ }\textbf {\bibinfo
  {volume} {84}},\ \bibinfo {pages} {4252} (\bibinfo {year}
  {2000})}\BibitemShut {NoStop}%
\bibitem [{\citenamefont {Gonoskov}\ \emph {et~al.}(2016)\citenamefont
  {Gonoskov}, \citenamefont {Tsatrafyllis}, \citenamefont {Kominis},\ and\
  \citenamefont {Tzallas}}]{Gonos16}%
  \BibitemOpen
  \bibfield  {author} {\bibinfo {author} {\bibfnamefont {I.~A.}\ \bibnamefont
  {Gonoskov}}, \bibinfo {author} {\bibfnamefont {N.}~\bibnamefont
  {Tsatrafyllis}}, \bibinfo {author} {\bibfnamefont {I.~K.}\ \bibnamefont
  {Kominis}},\ and\ \bibinfo {author} {\bibfnamefont {P.}~\bibnamefont
  {Tzallas}},\ }\bibfield  {title} {\bibinfo {title} {Quantum optical
  signatures in strong-field laser physics: Infrared photon counting in
  high-order-harmonic generation},\ }\href@noop {} {\bibfield  {journal}
  {\bibinfo  {journal} {Sci. Rep.}\ }\textbf {\bibinfo {volume} {6}},\ \bibinfo
  {pages} {32821} (\bibinfo {year} {2016})}\BibitemShut {NoStop}%
\bibitem [{\citenamefont {Gombk\"ot\H{o}}\ \emph {et~al.}(2016)\citenamefont
  {Gombk\"ot\H{o}}, \citenamefont {Czirj\'ak}, \citenamefont {Varr\'o},\ and\
  \citenamefont {F\"oldi}}]{GCVF16}%
  \BibitemOpen
  \bibfield  {author} {\bibinfo {author} {\bibfnamefont {A.}~\bibnamefont
  {Gombk\"ot\H{o}}}, \bibinfo {author} {\bibfnamefont {A.}~\bibnamefont
  {Czirj\'ak}}, \bibinfo {author} {\bibfnamefont {S.}~\bibnamefont {Varr\'o}},\
  and\ \bibinfo {author} {\bibfnamefont {P.}~\bibnamefont {F\"oldi}},\
  }\bibfield  {title} {\bibinfo {title} {Quantum-optical model for the dynamics
  of high-order-harmonic generation},\ }\href@noop {} {\bibfield  {journal}
  {\bibinfo  {journal} {Phys. Rev. A}\ }\textbf {\bibinfo {volume} {94}},\
  \bibinfo {pages} {013853} (\bibinfo {year} {2016})}\BibitemShut {NoStop}%
\bibitem [{\citenamefont {Gombk\"{o}t\H{o}}\ \emph {et~al.}(2021)\citenamefont
  {Gombk\"{o}t\H{o}}, \citenamefont {F\"{o}ldi},\ and\ \citenamefont
  {Varr\'o}}]{AGPF21}%
  \BibitemOpen
  \bibfield  {author} {\bibinfo {author} {\bibfnamefont {A.}~\bibnamefont
  {Gombk\"{o}t\H{o}}}, \bibinfo {author} {\bibfnamefont {P.}~\bibnamefont
  {F\"{o}ldi}},\ and\ \bibinfo {author} {\bibfnamefont {S.}~\bibnamefont
  {Varr\'o}},\ }\bibfield  {title} {\bibinfo {title} {A quantum optical
  description of photon statistics and cross-correlations in high harmonic
  generation},\ }\href@noop {} {\bibfield  {journal} {\bibinfo  {journal}
  {Phys. Rev. A}\ }\textbf {\bibinfo {volume} {104}},\ \bibinfo {pages}
  {033703} (\bibinfo {year} {2021})}\BibitemShut {NoStop}%
\bibitem [{\citenamefont {F\"{o}ldi}\ \emph {et~al.}(2021)\citenamefont
  {F\"{o}ldi}, \citenamefont {Magashegyi}, \citenamefont {Gombk\"{o}t\H{o}},\
  and\ \citenamefont {Varr\'o}}]{photonics8070263}%
  \BibitemOpen
  \bibfield  {author} {\bibinfo {author} {\bibfnamefont {P.}~\bibnamefont
  {F\"{o}ldi}}, \bibinfo {author} {\bibfnamefont {I.}~\bibnamefont
  {Magashegyi}}, \bibinfo {author} {\bibfnamefont {A.}~\bibnamefont
  {Gombk\"{o}t\H{o}}},\ and\ \bibinfo {author} {\bibfnamefont {S.}~\bibnamefont
  {Varr\'o}},\ }\bibfield  {title} {\bibinfo {title} {Describing high-order
  harmonic generation using quantum optical models},\ }\href@noop {} {\bibfield
   {journal} {\bibinfo  {journal} {Photonics}\ }\textbf {\bibinfo {volume} {8}}
  (\bibinfo {year} {2021})}\BibitemShut {NoStop}%
\bibitem [{\citenamefont {\'{A}kos Gombk\"{o}t\H{o}}(2023)}]{GA23}%
  \BibitemOpen
  \bibfield  {author} {\bibinfo {author} {\bibnamefont {\'{A}kos
  Gombk\"{o}t\H{o}}},\ }\bibfield  {title} {\bibinfo {title} {Squeezing
  properties of degenerate high-order hyper-raman lines emitted by a two-level
  system},\ }\href@noop {} {\bibfield  {journal} {\bibinfo  {journal} {Acta
  Physica Polonica Series a}\ }\textbf {\bibinfo {volume} {143}} (\bibinfo
  {year} {2023})}\BibitemShut {NoStop}%
\bibitem [{\citenamefont {Gombk\"ot{\H{o}}}\ \emph {et~al.}(2020)\citenamefont
  {Gombk\"ot{\H{o}}}, \citenamefont {Varr\'o}, \citenamefont {Mati},\ and\
  \citenamefont {F\"oldi}}]{GVMF20}%
  \BibitemOpen
  \bibfield  {author} {\bibinfo {author} {\bibfnamefont {A.}~\bibnamefont
  {Gombk\"ot{\H{o}}}}, \bibinfo {author} {\bibfnamefont {S.}~\bibnamefont
  {Varr\'o}}, \bibinfo {author} {\bibfnamefont {P.}~\bibnamefont {Mati}},\ and\
  \bibinfo {author} {\bibfnamefont {P.}~\bibnamefont {F\"oldi}},\ }\bibfield
  {title} {\bibinfo {title} {High-order harmonic generation as induced by a
  quantized field: Phase-space picture},\ }\href@noop {} {\bibfield  {journal}
  {\bibinfo  {journal} {Phys. Rev. A}\ }\textbf {\bibinfo {volume} {101}},\
  \bibinfo {pages} {013418} (\bibinfo {year} {2020})}\BibitemShut {NoStop}%
\bibitem [{\citenamefont {Gorlach}\ \emph {et~al.}(2020)\citenamefont
  {Gorlach}, \citenamefont {Neufeld}, \citenamefont {Rivera}, \citenamefont
  {Cohen},\ and\ \citenamefont {Kaminer}}]{Gorlach20}%
  \BibitemOpen
  \bibfield  {author} {\bibinfo {author} {\bibfnamefont {A.}~\bibnamefont
  {Gorlach}}, \bibinfo {author} {\bibfnamefont {O.}~\bibnamefont {Neufeld}},
  \bibinfo {author} {\bibfnamefont {N.}~\bibnamefont {Rivera}}, \bibinfo
  {author} {\bibfnamefont {O.}~\bibnamefont {Cohen}},\ and\ \bibinfo {author}
  {\bibfnamefont {I.}~\bibnamefont {Kaminer}},\ }\bibfield  {title} {\bibinfo
  {title} {The quantum-optical nature of high harmonic generation},\
  }\href@noop {} {\bibfield  {journal} {\bibinfo  {journal} {Nat. Comm.}\
  }\textbf {\bibinfo {volume} {11}},\ \bibinfo {pages} {4598} (\bibinfo {year}
  {2020})}\BibitemShut {NoStop}%
\bibitem [{\citenamefont {Gorlach}\ \emph {et~al.}(2023)\citenamefont
  {Gorlach}, \citenamefont {Tzur}, \citenamefont {Birk}, \citenamefont
  {Kr\"uger}, \citenamefont {Rivera}, \citenamefont {Cohen},\ and\
  \citenamefont {Kaminer}}]{Gorlach23}%
  \BibitemOpen
  \bibfield  {author} {\bibinfo {author} {\bibfnamefont {A.}~\bibnamefont
  {Gorlach}}, \bibinfo {author} {\bibfnamefont {M.~E.}\ \bibnamefont {Tzur}},
  \bibinfo {author} {\bibfnamefont {M.}~\bibnamefont {Birk}}, \bibinfo {author}
  {\bibfnamefont {M.}~\bibnamefont {Kr\"uger}}, \bibinfo {author}
  {\bibfnamefont {N.}~\bibnamefont {Rivera}}, \bibinfo {author} {\bibfnamefont
  {O.}~\bibnamefont {Cohen}},\ and\ \bibinfo {author} {\bibfnamefont
  {I.}~\bibnamefont {Kaminer}},\ }\bibfield  {title} {\bibinfo {title}
  {High-harmonic generation driven by quantum light},\ }\bibfield  {journal}
  {\bibinfo  {journal} {Nat. Phys.}\ }\href
  {https://doi.org/10.1038/s41567-023-02127-y} {10.1038/s41567-023-02127-y}
  (\bibinfo {year} {2023})\BibitemShut {NoStop}%
\bibitem [{\citenamefont {Stammer}\ \emph {et~al.}(2023)\citenamefont
  {Stammer}, \citenamefont {Rivera-Dean}, \citenamefont {Maxwell},
  \citenamefont {Lamprou}, \citenamefont {Ord\'o\~nez}, \citenamefont
  {Ciappina}, \citenamefont {Tzallas},\ and\ \citenamefont
  {Lewenstein}}]{LewPRX23}%
  \BibitemOpen
  \bibfield  {author} {\bibinfo {author} {\bibfnamefont {P.}~\bibnamefont
  {Stammer}}, \bibinfo {author} {\bibfnamefont {J.}~\bibnamefont
  {Rivera-Dean}}, \bibinfo {author} {\bibfnamefont {A.}~\bibnamefont
  {Maxwell}}, \bibinfo {author} {\bibfnamefont {T.}~\bibnamefont {Lamprou}},
  \bibinfo {author} {\bibfnamefont {A.}~\bibnamefont {Ord\'o\~nez}}, \bibinfo
  {author} {\bibfnamefont {M.~F.}\ \bibnamefont {Ciappina}}, \bibinfo {author}
  {\bibfnamefont {P.}~\bibnamefont {Tzallas}},\ and\ \bibinfo {author}
  {\bibfnamefont {M.}~\bibnamefont {Lewenstein}},\ }\bibfield  {title}
  {\bibinfo {title} {Quantum electrodynamics of intense laser-matter
  interactions: A tool for quantum state engineering},\ }\href@noop {}
  {\bibfield  {journal} {\bibinfo  {journal} {PRX Quantum}\ }\textbf {\bibinfo
  {volume} {4}},\ \bibinfo {pages} {010201} (\bibinfo {year}
  {2023})}\BibitemShut {NoStop}%
\bibitem [{\citenamefont {Walls}\ and\ \citenamefont {Tindle}(1975)}]{WT71}%
  \BibitemOpen
  \bibfield  {author} {\bibinfo {author} {\bibfnamefont {D.~F.}\ \bibnamefont
  {Walls}}\ and\ \bibinfo {author} {\bibfnamefont {C.~T.}\ \bibnamefont
  {Tindle}},\ }\bibfield  {title} {\bibinfo {title} {Nonlinear quantum effects
  in second-harmonic generation},\ }\href@noop {} {\bibfield  {journal}
  {\bibinfo  {journal} {Lettere al Nuovo Cimento}\ }\textbf {\bibinfo {volume}
  {2}},\ \bibinfo {pages} {915} (\bibinfo {year} {1975})}\BibitemShut {NoStop}%
\bibitem [{\citenamefont {Kwiat}\ \emph {et~al.}(1995)\citenamefont {Kwiat},
  \citenamefont {Mattle}, \citenamefont {Weinfurter}, \citenamefont
  {Zeilinger}, \citenamefont {Sergienko},\ and\ \citenamefont
  {Shih}}]{Kwiat95}%
  \BibitemOpen
  \bibfield  {author} {\bibinfo {author} {\bibfnamefont {P.~G.}\ \bibnamefont
  {Kwiat}}, \bibinfo {author} {\bibfnamefont {K.}~\bibnamefont {Mattle}},
  \bibinfo {author} {\bibfnamefont {H.}~\bibnamefont {Weinfurter}}, \bibinfo
  {author} {\bibfnamefont {A.}~\bibnamefont {Zeilinger}}, \bibinfo {author}
  {\bibfnamefont {A.~V.}\ \bibnamefont {Sergienko}},\ and\ \bibinfo {author}
  {\bibfnamefont {Y.}~\bibnamefont {Shih}},\ }\bibfield  {title} {\bibinfo
  {title} {New high-intensity source of polarization-entangled photon pairs},\
  }\href@noop {} {\bibfield  {journal} {\bibinfo  {journal} {Phys. Rev. Lett.}\
  }\textbf {\bibinfo {volume} {75}},\ \bibinfo {pages} {4337} (\bibinfo {year}
  {1995})}\BibitemShut {NoStop}%
\bibitem [{\citenamefont {Hochrainer}\ \emph {et~al.}(2022)\citenamefont
  {Hochrainer}, \citenamefont {Lahiri}, \citenamefont {Erhard}, \citenamefont
  {Krenn},\ and\ \citenamefont {Zeilinger}}]{Z22}%
  \BibitemOpen
  \bibfield  {author} {\bibinfo {author} {\bibfnamefont {A.}~\bibnamefont
  {Hochrainer}}, \bibinfo {author} {\bibfnamefont {M.}~\bibnamefont {Lahiri}},
  \bibinfo {author} {\bibfnamefont {M.}~\bibnamefont {Erhard}}, \bibinfo
  {author} {\bibfnamefont {M.}~\bibnamefont {Krenn}},\ and\ \bibinfo {author}
  {\bibfnamefont {A.}~\bibnamefont {Zeilinger}},\ }\bibfield  {title} {\bibinfo
  {title} {Quantum indistinguishability by path identity and with undetected
  photons},\ }\href@noop {} {\bibfield  {journal} {\bibinfo  {journal} {Rev.
  Mod. Phys.}\ }\textbf {\bibinfo {volume} {94}},\ \bibinfo {pages} {025007}
  (\bibinfo {year} {2022})}\BibitemShut {NoStop}%
\bibitem [{\citenamefont {Walls}\ and\ \citenamefont {Milburn}(1994)}]{WM94}%
  \BibitemOpen
  \bibfield  {author} {\bibinfo {author} {\bibfnamefont {D.~F.}\ \bibnamefont
  {Walls}}\ and\ \bibinfo {author} {\bibfnamefont {G.~J.}\ \bibnamefont
  {Milburn}},\ }\href@noop {} {\emph {\bibinfo {title} {Quantum Optics}}}\
  (\bibinfo  {publisher} {Springer-Verlag},\ \bibinfo {address} {Berlin},\
  \bibinfo {year} {1994})\BibitemShut {NoStop}%
\bibitem [{\citenamefont {Hillery}\ \emph {et~al.}(1994)\citenamefont
  {Hillery}, \citenamefont {Yu},\ and\ \citenamefont {Bergou}}]{HDB94}%
  \BibitemOpen
  \bibfield  {author} {\bibinfo {author} {\bibfnamefont {M.}~\bibnamefont
  {Hillery}}, \bibinfo {author} {\bibfnamefont {D.}~\bibnamefont {Yu}},\ and\
  \bibinfo {author} {\bibfnamefont {J.}~\bibnamefont {Bergou}},\ }\bibfield
  {title} {\bibinfo {title} {Effect of the pump state on the behavior of the
  degenerate parametric amplifier},\ }\href@noop {} {\bibfield  {journal}
  {\bibinfo  {journal} {Phys. Rev. A}\ }\textbf {\bibinfo {volume} {49}},\
  \bibinfo {pages} {1288} (\bibinfo {year} {1994})}\BibitemShut {NoStop}%
\bibitem [{\citenamefont {Louisell}\ \emph {et~al.}(1961)\citenamefont
  {Louisell}, \citenamefont {Yariv},\ and\ \citenamefont {Siegman}}]{L61}%
  \BibitemOpen
  \bibfield  {author} {\bibinfo {author} {\bibfnamefont {W.~H.}\ \bibnamefont
  {Louisell}}, \bibinfo {author} {\bibfnamefont {A.}~\bibnamefont {Yariv}},\
  and\ \bibinfo {author} {\bibfnamefont {A.~E.}\ \bibnamefont {Siegman}},\
  }\bibfield  {title} {\bibinfo {title} {Quantum fluctuations and noise in
  parametric processes. I},\ }\href@noop {} {\bibfield  {journal} {\bibinfo
  {journal} {Phys. Rev.}\ }\textbf {\bibinfo {volume} {124}},\ \bibinfo {pages}
  {1646} (\bibinfo {year} {1961})}\BibitemShut {NoStop}%
\bibitem [{\citenamefont {Gordon}\ \emph {et~al.}(1963)\citenamefont {Gordon},
  \citenamefont {Louisell},\ and\ \citenamefont {Walker}}]{G63}%
  \BibitemOpen
  \bibfield  {author} {\bibinfo {author} {\bibfnamefont {J.~P.}\ \bibnamefont
  {Gordon}}, \bibinfo {author} {\bibfnamefont {W.~H.}\ \bibnamefont
  {Louisell}},\ and\ \bibinfo {author} {\bibfnamefont {L.~R.}\ \bibnamefont
  {Walker}},\ }\bibfield  {title} {\bibinfo {title} {Quantum fluctuations and
  noise in parametric processes. II},\ }\href@noop {} {\bibfield  {journal}
  {\bibinfo  {journal} {Phys. Rev.}\ }\textbf {\bibinfo {volume} {129}},\
  \bibinfo {pages} {481} (\bibinfo {year} {1963})}\BibitemShut {NoStop}%
\bibitem [{\citenamefont {Hentschel}\ \emph {et~al.}(2001)\citenamefont
  {Hentschel}, \citenamefont {Kienberger}, \citenamefont {Spielmann},
  \citenamefont {Reider}, \citenamefont {Milosevic}, \citenamefont {Brabec},
  \citenamefont {Corkum}, \citenamefont {Heinzmann}, \citenamefont {Drescher},\
  and\ \citenamefont {Krausz}}]{Krausz_Nobel}%
  \BibitemOpen
  \bibfield  {author} {\bibinfo {author} {\bibfnamefont {M.}~\bibnamefont
  {Hentschel}}, \bibinfo {author} {\bibfnamefont {R.}~\bibnamefont
  {Kienberger}}, \bibinfo {author} {\bibfnamefont {C.}~\bibnamefont
  {Spielmann}}, \bibinfo {author} {\bibfnamefont {G.}~\bibnamefont {Reider}},
  \bibinfo {author} {\bibfnamefont {N.}~\bibnamefont {Milosevic}}, \bibinfo
  {author} {\bibfnamefont {T.}~\bibnamefont {Brabec}}, \bibinfo {author}
  {\bibfnamefont {P.}~\bibnamefont {Corkum}}, \bibinfo {author} {\bibfnamefont
  {U.}~\bibnamefont {Heinzmann}}, \bibinfo {author} {\bibfnamefont
  {M.}~\bibnamefont {Drescher}},\ and\ \bibinfo {author} {\bibfnamefont
  {F.}~\bibnamefont {Krausz}},\ }\bibfield  {title} {\bibinfo {title}
  {Attosecond metrology},\ }\href@noop {} {\bibfield  {journal} {\bibinfo
  {journal} {Nature}\ }\textbf {\bibinfo {volume} {414}},\ \bibinfo {pages}
  {509} (\bibinfo {year} {2001})}\BibitemShut {NoStop}%
\bibitem [{\citenamefont {Paul}\ \emph {et~al.}(2001)\citenamefont {Paul},
  \citenamefont {Toma}, \citenamefont {Breger}, \citenamefont {Mullot},
  \citenamefont {Augé}, \citenamefont {Balcou}, \citenamefont {Muller},\ and\
  \citenamefont {Agostini}}]{Agostini_Nobel}%
  \BibitemOpen
  \bibfield  {author} {\bibinfo {author} {\bibfnamefont {P.~M.}\ \bibnamefont
  {Paul}}, \bibinfo {author} {\bibfnamefont {E.~S.}\ \bibnamefont {Toma}},
  \bibinfo {author} {\bibfnamefont {P.}~\bibnamefont {Breger}}, \bibinfo
  {author} {\bibfnamefont {G.}~\bibnamefont {Mullot}}, \bibinfo {author}
  {\bibfnamefont {F.}~\bibnamefont {Augé}}, \bibinfo {author} {\bibfnamefont
  {P.}~\bibnamefont {Balcou}}, \bibinfo {author} {\bibfnamefont {H.~G.}\
  \bibnamefont {Muller}},\ and\ \bibinfo {author} {\bibfnamefont
  {P.}~\bibnamefont {Agostini}},\ }\bibfield  {title} {\bibinfo {title}
  {Observation of a train of attosecond pulses from high harmonic generation},\
  }\href@noop {} {\bibfield  {journal} {\bibinfo  {journal} {Science}\ }\textbf
  {\bibinfo {volume} {292}},\ \bibinfo {pages} {1689} (\bibinfo {year}
  {2001})}\BibitemShut {NoStop}%
\bibitem [{\citenamefont {Hillery}(1990)}]{H90}%
  \BibitemOpen
  \bibfield  {author} {\bibinfo {author} {\bibfnamefont {M.}~\bibnamefont
  {Hillery}},\ }\bibfield  {title} {\bibinfo {title} {Photon number divergence
  in the quantum theory of n-photon down conversion},\ }\href@noop {}
  {\bibfield  {journal} {\bibinfo  {journal} {Phys. Rev. A}\ }\textbf {\bibinfo
  {volume} {42}},\ \bibinfo {pages} {498} (\bibinfo {year} {1990})}\BibitemShut
  {NoStop}%
\bibitem [{\citenamefont {Schleich}(2001)}]{S01}%
  \BibitemOpen
  \bibfield  {author} {\bibinfo {author} {\bibfnamefont {W.~P.}\ \bibnamefont
  {Schleich}},\ }\href@noop {} {\emph {\bibinfo {title} {Quantum Optics in
  Phase Space}}}\ (\bibinfo  {publisher} {Wiley-VCH},\ \bibinfo {address}
  {Berlin},\ \bibinfo {year} {2001})\BibitemShut {NoStop}%
\bibitem [{\citenamefont {Benedict}\ and\ \citenamefont
  {Czirj\'ak}(1995)}]{BC95}%
  \BibitemOpen
  \bibfield  {author} {\bibinfo {author} {\bibfnamefont {M.~G.}\ \bibnamefont
  {Benedict}}\ and\ \bibinfo {author} {\bibfnamefont {A.}~\bibnamefont
  {Czirj\'ak}},\ }\bibfield  {title} {\bibinfo {title} {Generalized parity and
  quasi-probability density functions},\ }\href@noop {} {\bibfield  {journal}
  {\bibinfo  {journal} {J. Phys. A: Math. Gen.}\ }\textbf {\bibinfo {volume}
  {28}},\ \bibinfo {pages} {4599} (\bibinfo {year} {1995})}\BibitemShut
  {NoStop}%
\bibitem [{\citenamefont {Czirják}\ and\ \citenamefont
  {Benedict}(1996)}]{Czirjak_1996}%
  \BibitemOpen
  \bibfield  {author} {\bibinfo {author} {\bibfnamefont {A.}~\bibnamefont
  {Czirják}}\ and\ \bibinfo {author} {\bibfnamefont {M.~G.}\ \bibnamefont
  {Benedict}},\ }\bibfield  {title} {\bibinfo {title} {Joint Wigner function
  for atom - field interactions},\ }\href@noop {} {\bibfield  {journal}
  {\bibinfo  {journal} {Quantum and Semiclassical Optics: Journal of the
  European Optical Society Part B}\ }\textbf {\bibinfo {volume} {8}},\ \bibinfo
  {pages} {975} (\bibinfo {year} {1996})}\BibitemShut {NoStop}%
\bibitem [{\citenamefont {Mollow}\ and\ \citenamefont
  {Glauber}(1967{\natexlab{a}})}]{MG67I}%
  \BibitemOpen
  \bibfield  {author} {\bibinfo {author} {\bibfnamefont {B.~R.}\ \bibnamefont
  {Mollow}}\ and\ \bibinfo {author} {\bibfnamefont {R.~J.}\ \bibnamefont
  {Glauber}},\ }\bibfield  {title} {\bibinfo {title} {Quantum theory of
  parametric amplification. I},\ }\href@noop {} {\bibfield  {journal} {\bibinfo
   {journal} {Phys. Rev.}\ }\textbf {\bibinfo {volume} {160}},\ \bibinfo
  {pages} {1076} (\bibinfo {year} {1967}{\natexlab{a}})}\BibitemShut {NoStop}%
\bibitem [{\citenamefont {Mollow}\ and\ \citenamefont
  {Glauber}(1967{\natexlab{b}})}]{MG67II}%
  \BibitemOpen
  \bibfield  {author} {\bibinfo {author} {\bibfnamefont {B.~R.}\ \bibnamefont
  {Mollow}}\ and\ \bibinfo {author} {\bibfnamefont {R.~J.}\ \bibnamefont
  {Glauber}},\ }\bibfield  {title} {\bibinfo {title} {Quantum theory of
  parametric amplification. II},\ }\href@noop {} {\bibfield  {journal}
  {\bibinfo  {journal} {Phys. Rev.}\ }\textbf {\bibinfo {volume} {160}},\
  \bibinfo {pages} {1097} (\bibinfo {year} {1967}{\natexlab{b}})}\BibitemShut
  {NoStop}%
\bibitem [{\citenamefont {Drobn\'y}\ and\ \citenamefont {Jex}(1992)}]{DJ92}%
  \BibitemOpen
  \bibfield  {author} {\bibinfo {author} {\bibfnamefont {G.}~\bibnamefont
  {Drobn\'y}}\ and\ \bibinfo {author} {\bibfnamefont {I.}~\bibnamefont {Jex}},\
  }\bibfield  {title} {\bibinfo {title} {Collapses and revivals in the energy
  exchange in the process of k-photon down-conversion with quantized pump},\
  }\href@noop {} {\bibfield  {journal} {\bibinfo  {journal} {Phys. Rev. A}\
  }\textbf {\bibinfo {volume} {45}},\ \bibinfo {pages} {1816} (\bibinfo {year}
  {1992})}\BibitemShut {NoStop}%
\bibitem [{\citenamefont {Bhattacharya}\ \emph {et~al.}(2023)\citenamefont
  {Bhattacharya}, \citenamefont {Lamprou}, \citenamefont {Maxwell},
  \citenamefont {Ordonez}, \citenamefont {Pisanty}, \citenamefont
  {Rivera-Dean}, \citenamefont {Stammer}, \citenamefont {Ciappina},
  \citenamefont {Lewenstein},\ and\ \citenamefont {Tzallas}}]{Tzallas_2023}%
  \BibitemOpen
  \bibfield  {author} {\bibinfo {author} {\bibfnamefont {U.}~\bibnamefont
  {Bhattacharya}}, \bibinfo {author} {\bibfnamefont {T.}~\bibnamefont
  {Lamprou}}, \bibinfo {author} {\bibfnamefont {A.~S.}\ \bibnamefont
  {Maxwell}}, \bibinfo {author} {\bibfnamefont {A.}~\bibnamefont {Ordonez}},
  \bibinfo {author} {\bibfnamefont {E.}~\bibnamefont {Pisanty}}, \bibinfo
  {author} {\bibfnamefont {J.}~\bibnamefont {Rivera-Dean}}, \bibinfo {author}
  {\bibfnamefont {P.}~\bibnamefont {Stammer}}, \bibinfo {author} {\bibfnamefont
  {M.~F.}\ \bibnamefont {Ciappina}}, \bibinfo {author} {\bibfnamefont
  {M.}~\bibnamefont {Lewenstein}},\ and\ \bibinfo {author} {\bibfnamefont
  {P.}~\bibnamefont {Tzallas}},\ }\bibfield  {title} {\bibinfo {title}
  {Strong-laser-field physics, non-classical light states and quantum
  information science},\ }\href@noop {} {\bibfield  {journal} {\bibinfo
  {journal} {Reports on Progress in Physics}\ }\textbf {\bibinfo {volume} {86}}
  (\bibinfo {year} {2023})}\BibitemShut {NoStop}%
\bibitem [{\citenamefont {Czirjak}\ \emph {et~al.}(2013)\citenamefont
  {Czirjak}, \citenamefont {Majorosi}, \citenamefont {Kovacs},\ and\
  \citenamefont {Benedict}}]{Czirjak_2013}%
  \BibitemOpen
  \bibfield  {author} {\bibinfo {author} {\bibfnamefont {A.}~\bibnamefont
  {Czirjak}}, \bibinfo {author} {\bibfnamefont {S.}~\bibnamefont {Majorosi}},
  \bibinfo {author} {\bibfnamefont {J.}~\bibnamefont {Kovacs}},\ and\ \bibinfo
  {author} {\bibfnamefont {M.~G.}\ \bibnamefont {Benedict}},\ }\bibfield
  {title} {\bibinfo {title} {Emergence of oscillations in quantum entanglement
  during rescattering},\ }\href@noop {} {\bibfield  {journal} {\bibinfo
  {journal} {Physica Scripta}\ }\textbf {\bibinfo {volume} {T153}} (\bibinfo
  {year} {2013})}\BibitemShut {NoStop}%
\bibitem [{\citenamefont {Majorosi}\ \emph {et~al.}(2017)\citenamefont
  {Majorosi}, \citenamefont {Benedict},\ and\ \citenamefont
  {Czirjak}}]{Majorosi_2017}%
  \BibitemOpen
  \bibfield  {author} {\bibinfo {author} {\bibfnamefont {S.}~\bibnamefont
  {Majorosi}}, \bibinfo {author} {\bibfnamefont {M.~G.}\ \bibnamefont
  {Benedict}},\ and\ \bibinfo {author} {\bibfnamefont {A.}~\bibnamefont
  {Czirjak}},\ }\bibfield  {title} {\bibinfo {title} {Quantum entanglement in
  strong-field ionization},\ }\href@noop {} {\bibfield  {journal} {\bibinfo
  {journal} {Phys. Rev. A}\ }\textbf {\bibinfo {volume} {96}} (\bibinfo {year}
  {2017})}\BibitemShut {NoStop}%
\bibitem [{\citenamefont {Maxwell}\ \emph {et~al.}(2022)\citenamefont
  {Maxwell}, \citenamefont {Madsen},\ and\ \citenamefont
  {Lewenstein}}]{Lewenstein_2022}%
  \BibitemOpen
  \bibfield  {author} {\bibinfo {author} {\bibfnamefont {A.~S.}\ \bibnamefont
  {Maxwell}}, \bibinfo {author} {\bibfnamefont {L.~B.}\ \bibnamefont
  {Madsen}},\ and\ \bibinfo {author} {\bibfnamefont {M.}~\bibnamefont
  {Lewenstein}},\ }\bibfield  {title} {\bibinfo {title} {Entanglement of
  orbital angular momentum in non-sequential double ionization},\ }\href@noop
  {} {\bibfield  {journal} {\bibinfo  {journal} {Nat. Comm.}\ }\textbf
  {\bibinfo {volume} {13}} (\bibinfo {year} {2022})}\BibitemShut {NoStop}%
\bibitem [{\citenamefont {Appi}\ \emph {et~al.}(2023)\citenamefont {Appi},
  \citenamefont {Weissenbilder}, \citenamefont {Nagyill\'{e}s}, \citenamefont
  {Diveki}, \citenamefont {Peschel}, \citenamefont {Farkas}, \citenamefont
  {Plach}, \citenamefont {Vismarra}, \citenamefont {Poulain}, \citenamefont
  {Weber}, \citenamefont {Arnold}, \citenamefont {Varj\'{u}}, \citenamefont
  {Kahaly}, \citenamefont {Eng-Johnsson},\ and\ \citenamefont
  {L'Huillier}}]{Appi23}%
  \BibitemOpen
  \bibfield  {author} {\bibinfo {author} {\bibfnamefont {E.}~\bibnamefont
  {Appi}}, \bibinfo {author} {\bibfnamefont {R.}~\bibnamefont {Weissenbilder}},
  \bibinfo {author} {\bibfnamefont {B.}~\bibnamefont {Nagyill\'{e}s}}, \bibinfo
  {author} {\bibfnamefont {Z.}~\bibnamefont {Diveki}}, \bibinfo {author}
  {\bibfnamefont {J.}~\bibnamefont {Peschel}}, \bibinfo {author} {\bibfnamefont
  {B.}~\bibnamefont {Farkas}}, \bibinfo {author} {\bibfnamefont
  {M.}~\bibnamefont {Plach}}, \bibinfo {author} {\bibfnamefont
  {F.}~\bibnamefont {Vismarra}}, \bibinfo {author} {\bibfnamefont
  {V.}~\bibnamefont {Poulain}}, \bibinfo {author} {\bibfnamefont
  {N.}~\bibnamefont {Weber}}, \bibinfo {author} {\bibfnamefont {C.~L.}\
  \bibnamefont {Arnold}}, \bibinfo {author} {\bibfnamefont {K.}~\bibnamefont
  {Varj\'{u}}}, \bibinfo {author} {\bibfnamefont {S.}~\bibnamefont {Kahaly}},
  \bibinfo {author} {\bibfnamefont {P.}~\bibnamefont {Eng-Johnsson}},\ and\
  \bibinfo {author} {\bibfnamefont {A.}~\bibnamefont {L'Huillier}},\ }\bibfield
   {title} {\bibinfo {title} {Two phase-matching regimes in high-order harmonic
  generation},\ }\href@noop {} {\bibfield  {journal} {\bibinfo  {journal} {Opt.
  Express}\ }\textbf {\bibinfo {volume} {31}},\ \bibinfo {pages} {31687}
  (\bibinfo {year} {2023})}\BibitemShut {NoStop}%
\end{thebibliography}

%

\end{document}